\documentclass[sigconf,nonacm]{acmart}
\pdfoutput=1

\newcommand{\repourl}{https://github.com/Visual-Computing/DynamicExplorationGraph/tree/crEG}

\makeatletter
  \if@ACM@anonymous
    \renewcommand{\repourl}{https://anonymous.4open.science/r/B8F8} 
  \fi
\makeatother

\usepackage{graphicx}
\usepackage[list=true, font=small, labelfont=bf, 
labelformat=brace, position=top]{subcaption}

\usepackage{tabularx}
\usepackage{makecell}
\usepackage{multirow}

\usepackage{float}
\usepackage{enumitem}
\usepackage{footmisc}

\usepackage{algorithm} 
\usepackage[noend]{algpseudocode} 
\usepackage{amsmath,amsfonts}
\usepackage{bm} 

\usepackage{balance}

\setlist[description]{leftmargin=\parindent,}

\newcommand{\Break}{\State \textbf{break}}

\DeclareMathOperator{\argmin}{argmin}
\DeclareMathOperator{\argmax}{argmax} 

\DeclareMathOperator{\gain}{gain}
\DeclareMathOperator{\ANNS}{ANNS}
\DeclareMathOperator{\KNN}{KNNS}
\DeclareMathOperator{\GQ}{GQ}

\DeclareMathOperator{\AND}{\overline{\delta_N}}
\DeclareMathOperator{\crEG}{crEG}

\DeclareMathOperator{\va}{\large \text{v}_A}
\DeclareMathOperator{\vb}{\large \text{v}_B}
\DeclareMathOperator{\vc}{\large \text{v}_C}
\DeclareMathOperator{\vd}{\large \text{v}_D}
\DeclareMathOperator{\ve}{\large \text{v}_E}
\DeclareMathOperator{\vf}{\large \text{v}_F}

\DeclareMathOperator{\opt}{opt}

\DeclareMathOperator{\extend}{ext}

\newcommand\recallAtVar[1]{\operatorname{recall@}\hspace{-0.2em}{#1}}

\AtBeginDocument{%
  \fancypagestyle{firstpagestyle}{%
    \fancyfoot[L]{\scriptsize %
      \copyright\ ACM 2024. This is the author's version of the work. It is posted here for your personal use. Not for redistribution.\\%
      The definitive Version of Record was published in \textit{Proceedings of the 2024 International Conference on Multimedia Retrieval (ICMR '24)}, %
      \url{https://doi.org/10.1145/3652583.3658117}.%
    }%
  }%
}

\acmConference[ICMR '24]{Proceedings of the 2024 International Conference on Multimedia Retrieval}{June 10--14, 2024}{Phuket, Thailand}
\acmBooktitle{Proceedings of the 2024 International Conference on Multimedia Retrieval (ICMR '24), June 10--14, 2024, Phuket, Thailand}
\acmISBN{979-8-4007-0619-6/24/06}
\acmDOI{10.1145/3652583.3658117}
\acmYear{2024}
\copyrightyear{2024}

\begin{document}

\title{An Exploration Graph with Continuous Refinement for Efficient Multimedia Retrieval}

\author{Nico Hezel}
\orcid{0000-0002-3957-4672}
\affiliation{%
  \institution{HTW Berlin}
  \city{Berlin}
  \country{Germany}
}
\email{hezel@htw-berlin.de}

\author{Kai Uwe Barthel}
\orcid{0000-0001-6309-572X}
\affiliation{%
  \institution{HTW Berlin}
  \city{Berlin}
  \country{Germany}
}
\email{barthel@htw-berlin.de}

\author{Konstantin Schall}
\orcid{0000-0003-3548-0537}
\affiliation{%
  \institution{HTW Berlin}
  \city{Berlin}
  \country{Germany}
}
\email{konstantin.schall@htw-berlin.de}

\author{Klaus Jung}
\orcid{0000-0002-3600-6848}
\affiliation{%
  \institution{HTW Berlin}
  \city{Berlin}
  \country{Germany}
}
\email{klaus.jung@htw-berlin.de}

%
%
%

\renewcommand{\shortauthors}{Nico Hezel, Kai Uwe Barthel, Konstantin Schall, \& Klaus Jung} 

\begin{abstract}
As datasets and the dimensionality of feature vectors continue to grow, Approximate Nearest Neighbor Search (ANNS) in large multimedia databases becomes increasingly relevant. Graph-based approaches have demonstrated to offer the best trade-off between retrieval precision and search time. Despite their ability to deliver search times several orders of magnitude faster than exact search techniques, existing methods suffer from slow constructions speeds or high memory requirements.
This paper presents a \textit{continuous refining Exploration Graph} (crEG), a novel approach for rapidly constructing a compact exploration graph with state-of-the-art search performance. Additionally, it provides the ability to enhance its effectiveness even further through an optional edge optimization algorithm. Both algorithms are specifically designed to produce and operate on undirected graphs with even degrees and guarantee graph connectivity at any time – a property particularly valuable for \textit{exploratory search}, where the query is part of the database elements.
Although such queries provide an advantageous starting point for graph search algorithms, they have been rarely considered in the context of ANNS, yet are crucial for recommendation and exploration systems. 
Our experiments demonstrate high efficiency in ANNS does not necessarily translate to a good performance in \textit{exploratory search}. 
\end{abstract}





\maketitle

\section{Introduction}

Nearest Neighbor Search (NNS) finds the closest data points to a query in a dataset, where proximity depends on a chosen similarity metric. The data points are commonly represented as dense feature vectors, often derived from the activation energy of a deep neural network layer \cite{Simonyan2015}. A linear search based on these features becomes less efficient as the number of dimensions or number of data points increases \cite{Li2020}. To address this problem, numerous applications \cite{Zhao2019,Johnson2019,Sugawara2016} resort to approximated nearest neighbor search (ANNS) \cite{Wei2020,Wang2021}, a slightly less accurate but much faster solution. A variation of ANNS is used in product recommender systems \cite{Park2015} and visual image browsing systems \cite{Barthel2019}, where the goal is to retrieve many items similar to a selected item from the dataset.

\subsection{Background}

\textbf{Exact vs. Approximate Search:} Due to the curse of dimensionality exact $k$-nearest neighbor search (KNNS) is very inefficient for modern datasets \cite{Indyk1998}. Approximate search algorithms (ANNS) on the other hand, compress the data points or leveraging additional data structures to skip a considerable portion of them. For a finite set of data points $P$ and a set of queries $Q$ the recall@k is used to measure the quality of an ANNS algorithm when retrieving the $k$ closest data points from $P$ for each query $q$. 
\begin{equation}
\recallAtVar{k} = \frac{1}{|Q|} \sum_{q \in Q} \frac{\lvert \ANNS(P, q) \cap \KNN(P, q) \rvert}{k}
\label{eq:recallAtK}
\end{equation}
When evaluating ANNS algorithms, the recall rate should be considered in relation to the search time (queries per second) \cite{Aumuller2020}. 
In recent studies \cite{Aumuller2020,Shimomura2021,DPG2020} proximity graphs have been shown to offer the best trade-off between search accuracy and efficiency.
\\
\\
\noindent \textbf{Proximity Graph.} Let $G = G(V, E)$ be a directed graph, where $V$ is the set of vertices and $E$ the set of edges. For every data point in $P$ exists a vertex in $V$. Furthermore let $(v,u) \in E$ denote an edge connecting $v \in V$ with $u \in V$. Edges are considered short or long if the distance of the incident vertices $\delta(u,v)$ is small or large, respectively. Let the neighbors $N(G, v)$ refer to the set of adjacent vertices of $v$ in $G$ and $|N(G, v)|$ be the \textit{out-degree} of the vertex. A commonly used graph metric for ANNS is the \textit{graph quality} \cite{Dong2011}, which compares the similarity of the neighborhood of a vertex to its true nearest vertices:
\begin{equation}
\label{eq:graph_quality}
\GQ(G) = \frac{1}{|V|} \sum_{v \in V} \frac{|N(G, v) \cap \KNN(V, v)|}{|N(G, v)|}
\end{equation}
\noindent where $|\KNN(V, v)| = |N(G, v)|$ for every $v$.
\\
\\
\noindent \textbf{Graph Search.} There are two commonly used graph search algorithms: range-search and greedy-search\cite{Wang2021Survey}. Although they are quite similar, the greedy-search does not use a range-search factor $\varepsilon$ and instead increases $k$ internally to explore more vertices. In both cases the final search result is stored in $R$ with $|R| = k$. The algorithms start at one or many seed vertices $S \subset V$ which are either predefined, randomly selected, or chosen in some other way. In each iteration, also called \textit{hop}, the element in $S$ closest to the query is removed from $S$ and its neighbors are analyzed (see Algorithm \ref{alg:rangeSearch}). The closest neighbors are added to $R$. An auxiliary set $C$ of checked vertices prevents duplicate vertex examinations. 
The goal is to maximize $\recallAtVar{k}$ while minimizing $\lvert C \rvert$. 
The efficiency of navigating the graph from one vertex to another ("navigation speed"), is the ratio of hops and $\lvert C \rvert$.

\begin{algorithm}\small
    \caption{RangeSearch($G, S, q, k, \varepsilon$)} \label{alg:rangeSearch}
	\begin{algorithmic}[1]
	\Require graph $G$, set of seed vertices $S \subset V$, query $q \in \mathbb{R}^m$, number of search results $k \in \mathbb{N}$, search range factor $\varepsilon \in \mathbb{R}^+$
    \Ensure $R$ is a list of vertices approximately closest to $q$
    \State $C \gets \emptyset$ \Comment{set of checked vertices}
    \State $R \gets S$ \Comment{result list}
    \State $r \gets \infty$ \Comment{search radius}
    \While {$S \neq \emptyset$}
        \State $s \gets \argmin_{x \in S} \delta(x, q)$ \Comment{find closest vertex to $q$ in $S$}
        \State $S \gets S \setminus \{s\}$
        \If{$\delta(s, q) > r(1 + \varepsilon)$} \Comment{best s is already too far way}
            \State \Return $R$ \Comment{stop \& return the current result list}
        \EndIf
        
        \State $N \gets N(G,s) \setminus C$ \Comment{unchecked neighbors of $s$}
        \ForAll {$n \in N$}
            \If{$\delta(n, q) \leq r(1 + \varepsilon)$} \Comment{only check close neighbors} 
                \State $S \gets S \cup \{n\}$
                \Comment{analyze their neighborhood later} 
                \If{$\delta(n, q) \leq r$} \Comment{is the neighbor close enough?} 
                    \State $R \gets R \cup \{n\}$
                    \Comment{add to the result list} 
                    \If{$|R| > k$} \Comment{limit the result list size}
                        \State $R \gets R \setminus \{ \argmax_{x \in R} \delta(x, q) \}$
                        \State $r \gets \max_{x \in R} \delta(x, q)$
                    \EndIf
                \EndIf
            \EndIf
        \EndFor
        \State $C \gets C \cup N$ \Comment{add neighbors to set of checked vertices}
    \EndWhile
    \State \Return $R$
	\end{algorithmic} 
\end{algorithm} 

\noindent \textbf{Problems of Graph-based Search Methods.} \textit{K-nearest neighbor graphs} like kGraph \cite{Dong2011} and EFANNA \cite{Cong2016} try to connect each vertex to the closest $k$ other vertices, which increases the \textit{graph quality} but makes navigation to other graph regions rather difficult and slow. Several other graphs (ONNG \cite{Iwasaki2018}, DPG \cite{DPG2020}, NSG \cite{Cong2019}, NSSG \cite{Cong2021}) prune the edges of an existing KNNG to improve its navigation efficiency. They require two graphs (pruned and non-pruned version) to be in memory at the same time and have a cumulative construction cost. 
Incremental graphs (NSW \cite{Malkov2014}, HNSW \cite{Malkov2020}) combine the acquisition of neighbor candidates and the pruning process in one step for each new vertex added to the graph. While the memory consumption is very low, even the fastest graph in this group, HNSW, is quite slow to construct. It also introduces a hierarchical data structure, which does not guarantee \textit{strong connectivity} on its lower layers, limiting \textit{exploratory searches}.
\\
\\
\noindent \textbf{Exploratory Search.} General exploration, where the query is an indexed data point in the graph and also the seed of the search, is rarely considered in ANNS literature. 
This form of search is used in various browsing applications, including interactive visual image navigation systems \cite{Navigu2023} and multi-label classification frameworks \cite{Hyvoenen2022}. 
While some graphs \cite{Dong2011,Cong2016} are inherently optimized for indexed queries, no experiments have been conducted with ideal search seeds.

\subsection{Contribution}
In this paper, we propose the \textit{continuous refining Exploration Graph (crEG)}, which consists of a single graph component and no additional data structures. There are no restrictions on the distance function and the feature space. 
The two manipulation algorithms for adding a new vertex and improving existing edges always maintain the regularity and connectivity of the graph, without compromising the search or exploration efficiency. 
\\
\\
Our main contributions are as follows:
\begin{itemize}
\item Introducing the fundamentals of the crEG (Section \ref{sec:graphFoundation}) and a new metric for assessing the quality of small graph changes which is extensively used by our manipulations algorithms.

\item A very fast graph construction strategy for even-regular, undirected, weighted graphs is presented in Section \ref{sec:incrementalConstruction}, which is several times faster than the current state of the art. 

\item Our edge optimization algorithm can improve an existing crEG even further, while it is being used in a production system. More details can be found in Section \ref{sec:continuousEdgeOptimizations}. 

\item Comprehensive experiments show that the search speed of crEG is up to 250\% higher than the current state of the art in approximate nearest neighbor search tasks (Section \ref{sec:searchExperiments}) .

\item A protocol for testing the exploration quality of graphs is established and used in Section \ref{sec:explorationExperiements}. The crEG is again up to 50\% more efficient for various recall ranges.

\end{itemize}

\section{Related Work}
Leading graph-based ANN methods approximate fundamental structures such as the \textit{k-Nearest Neighbor Graph} (KNNG) \cite{Paredes2005}, \textit{Relative Neighborhood Graph} (RNG) \cite{Toussaint1980}, or \textit{Delaunay Graph} (DG) \cite{Delaunay1933} because constructing these graphs is very slow and requires extensive knowledge of the data distribution and feature space \cite{Navarro2002}.

The authors of \textit{kGraph} \cite{Dong2011} build a KNNG by iteratively updating a random graph: identifying better neighbors in the broader neighborhood of each vertex and using the \textit{graph quality} to measure their success. EFANNA \cite{Cong2016}, following a similar approach, starts with multiple kd-trees \cite{Bentley1975} and searches them to build an initial KNNG, before refining the neighbors similar to \textit{kGraph}. However, the overall search efficiency of all KNNGs suffers from too many very similar neighbors and a resulting slow \textit{navigation speed} \cite{Peng2019}.

A \textit{Relative Neighborhood Graph} (RNG) or \textit{Delaunay Graph} (DG) mitigate this issue by constraining the edges of a vertex. In \cite{Malkov2014}, a \textit{Navigable Small World} (NSW) graph approximates such a \textit{Delaunay Graph} by incrementally identifying potential neighbor candidates for a new vertex through \textit{greedy search} and connecting them with undirected edges. As the number of vertices grows, the initially longer edges serve as shortcuts to various parts of the graph and a specific edge distribution forms. Since no edges are deleted or replaced, this process inevitably creates hub vertices with high out-degrees, leading to poly-logarithmic search complexity as the graph size increases \cite{Ponomarenko2014}. Later, the same authors proposed \textit{Hierarchical Navigable Small World} (HNSW) \cite{Malkov2020}, an approximation of a RNG by building a hierarchical graph and adding longer edges to higher layers for faster navigation. Additionally, an upper limit on the connections per vertex is imposed. Its search complexity scales logarithmically, but its hierarchical advantages decrease with higher intrinsic dimensionality \cite{Peng2019}. Moreover, the lower layers lack connectivity guarantees, potentially trapping the search process in local minima \cite{Cong2021}.

The \textit{Diversified Proximity Graph} (DPG) \cite{DPG2020} modifies an existing \textit{kGraph} by converting all edges into undirected edges. In a second step, edges to adjacent neighbors of a vertex are removed if their angular similarity is too high. This approximation of a RNG \cite{Wang2021Survey} evenly distributes the neighboring vertices in all directions and therefore increases the \textit{navigation speed}. However, depending on the initial \textit{kGraph}'s size, there is no guarantee that enough edges will be removed to create a manageable graph index. 

The concept of pruning the edges of an existing graph is used by many other graphs. For instance, the \textit{Optimized Nearest Neighbors Graph} (ONNG) \cite{Iwasaki2018} applies several in- and out-degree adjustments to approximate a \textit{Delaunay Graph} and \textit{Relative Neighborhood Graph}. The authors of \textit{Navigating Spreading-out Graph} (NSG) \cite{Cong2019} and \textit{Navigating Satellite System Graph} (NSSG) \cite{Cong2021} further reduce the number of edges by approximating a \textit{Monotonic Relative Neighborhood Graph} (MRNG). A MRNG never converges to a fully connected graph like DG \cite{Harwood2016} and guarantees a monotonic path for every pair of vertices ${p,q}$ such that the vertices in a path $v_1, v_2, ..., v_k$ with $v_1=p$ and $v_k=q$ come closer to $q$ with every step $\delta(v_i,q) > \delta(v_{i+1},q)$ for $i = 1,...,k - 1$. NSG and NSSG ensure connectivity because of a VP-tree which is used to combine potential graph components.

\section{Exploration Graph} 
\label{sec:graphConstructionAndOptimization}
To quickly build a proximity graph suitable for search and exploration, we propose the \textit{continuous refining Exploration Graph} (crEG). Its connectivity, a requirement for exploration, is obtained through incremental construction of an undirected graph \cite{Malkov2014}. A well-balanced distribution of neighboring vertices avoids the formation of hubs (vertices with high degrees) \cite{Wang2021Survey}. This distribution is guided by the graph's even regularity property and a low average neighbor distance. The crEG also approximates a Monotonic Relative Neighborhood Graph (MRNG) \cite{Cong2019} to reinforce the vertex distribution and aid construction.

\subsection{Exploration Graph Foundations} \label{sec:graphFoundation}
The following section presents the core principles of crEG: limiting edges per vertex, measuring the quality of neighborhood distribution, and proving connectivity under certain conditions.
\\
\\
\textbf{Even Regularity}. By design the crEG is a single even-regular, undirected graph without loops. Its regularity, $d$, makes the smallest possible $\crEG_d$ also a complete graph $K_{d+1}$ with $d+1$ vertices. 
The number of edges in any undirected regular graph is $|E| = (|V| \cdot d)/2$, derived by the handshaking lemma \cite{Euler1736}. The even regularity is vital as an odd regularity would require an even number of vertices in the graph at all times. The crEG requires a regularity value of 4 or higher in order to avoid circles and isolated vertices.
\\
\\
\textbf{Edge Quality}. When adding a new vertex, $d/2$ existing edges must be replaced by $d$ new edges to maintain regularity. To guide this process, we need to determine if a change improves the search efficiency. The following section will review the existing \textit{graph quality} metric (Equation \ref{eq:graph_quality}) and highlight its limitations. Subsequently, a new metric is proposed which is used by the crEG. 

\begin{figure}[h!]
\centering
\includegraphics[trim=4cm 6cm 3cm 6.5cm,clip=true,width=\columnwidth]{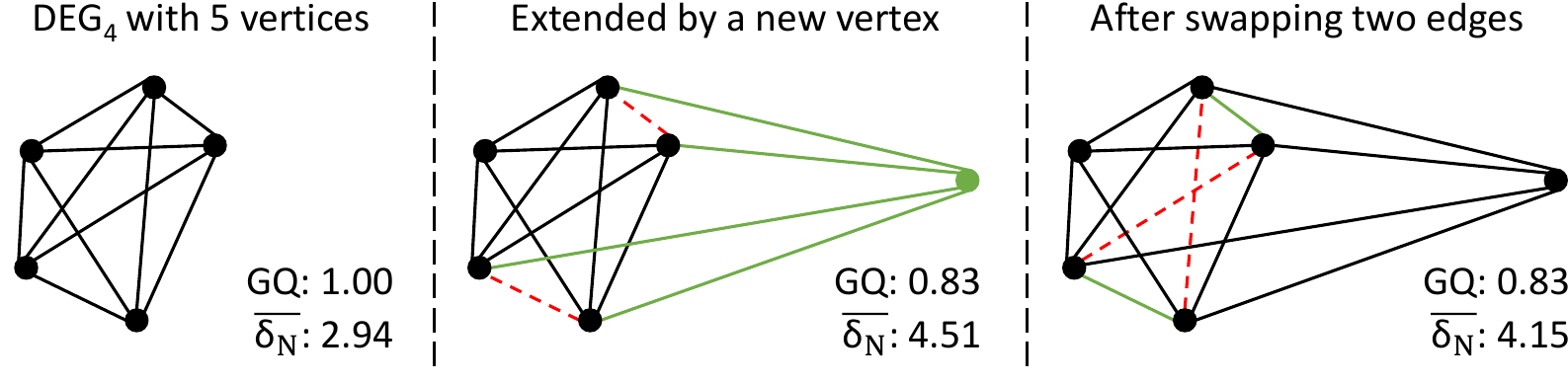} 
\vspace{-0.5cm}
\captionsetup{font=small}
\caption{A 2D toy example. Left: Smallest possible $crEG_{\bm{4}}$ ($\bm{K_5}$). Center: A new vertex is added. Right: Two edges have been swapped. Our new proposed metric $\AND$ recognizes the improvement while the graph quality (GQ) remains the same.
} \label{fig:deg_basics}
\end{figure}

Figure \ref{fig:deg_basics} on the left shows a $\crEG_4$ with 5 vertices which is equal to the complete graph $K_5$. 
A complete graph always has a perfect \textit{graph quality} (GQ) of 1, since all vertices are connected to each other.
In the center of Figure \ref{fig:deg_basics}, a new vertex (shown in green) is added to the graph.
To maintain regularity, two edges (red) were replaced by four new ones (green), which decreases the \textit{graph quality} in this case. 
On the right of Figure \ref{fig:deg_basics}, two existing edges have been swapped. 
Although the four involved vertices have been connected to closer vertices, the \textit{graph quality} remains the same.
Such insensitivity is often observed for small changes and makes clear decisions for graph construction impossible.
\\
We propose the \textbf{Average Neighbor Distance} ($\AND$) as a better metric to assess the quality of an undirected graph: Let $G(V,E)$ be a $d$-regular undirected graph, and $N(G, u) \subset V$ be the vertices adjacent to $u \in V$. The \textit{average neighbor distance} for a set of vertices $U \subset V$ is:
\begin{equation} \label{eq:averageNeighborDistance}
\AND(U) = \frac{1}{|U|} \sum_{u \in U} \frac{1}{d} \hspace{-0.1cm}\sum_{v \in N(G, u)} \delta(u, v)
\end{equation}
\\
In the case $U = V$, the average neighbor distance of the entire graph is calculated. 
A low distance indicates a high similarity between connected vertices. 
To reduce the computational effort, the distances of adjacent vertices are stored as their edge weights: $w_{u,v} = \delta(u, v)$. 
Changing an undirected edge $(u,v)$ in a crEG affects the \textit{average neighbor distance} of both vertices $u$ and $v$.
When swapping the endpoints of two edges $(u1,v1)$ and $(u2,v2)$, it is sufficient to compare the sum of the edge weights before ($w_{u1,v1} + w_{u2,v2}$) and after ($w_{u1, v2} + w_{u2, v1}$) the change to know if the \textit{average neighbor distance} of the graph is reduced. 
This efficient calculation is extensively used by all our graph manipulation algorithms.
Figure \ref{fig:deg_basics} illustrates how the metrics $\GQ$ and $\AND$ indicate a graph deterioration after adding a new vertex, but only $\AND$ was able to detect an improvement after two edges were swapped.
\\
\\
\textbf{Connectivity}. 
The crEG is also an Eulerian graph: each vertex has an even degree, guaranteeing a closed trail (Euler cycle) that includes all the graph's edges. Any vertex can serve as a starting point for this cycle. Consequently, every vertex has at least two paths to reach any other vertex. As the crEG lacks bridges, it guarantees 2-edge connectivity; we can remove one edge without disconnecting the graph. The Graph manipulation methods in Sections \ref{sec:incrementalConstruction} and \ref{sec:continuousEdgeOptimizations} are specifically designed to always preserve connectivity.

\subsection{Incremental Construction} 
\label{sec:incrementalConstruction}

\begin{figure}[h!]
\centering
\includegraphics[trim=5cm 3.5cm 6.2cm 3cm,clip=true,width=\columnwidth]{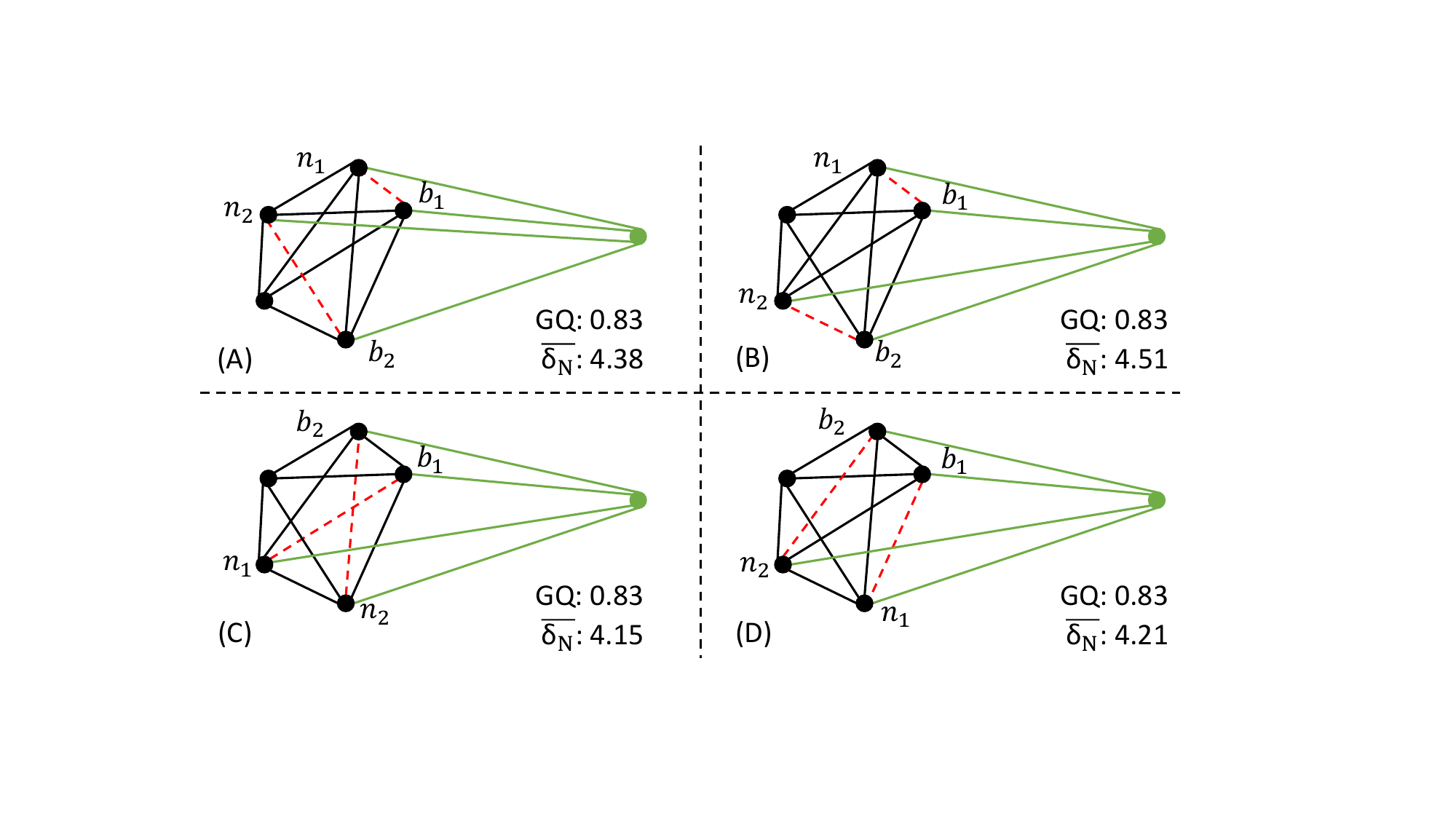}
\captionsetup{font=small}
\vspace{-0.5cm}
\caption{A 2D example for extending a $\crEG_4$, showing the final graphs for various selection schemes for choosing the neighbor $n$ in step (4): (A) the neighbor closest to $v$. (B) the neighbor with the shortest edge. (C) the neighbor with the longest edge. (D) the neighbor where the $\AND(V)$ decreases the most.} \label{fig:add_vertex}
\end{figure}

An existing crEG can be extended by connecting a new vertex $v$ to the approximated nearest vertices of a graph search. This process requires the removal of $d/2$ edges and the addition of $d$ new ones:

\begin{enumerate}
\item Starting from the smallest $\crEG_d$ with $d+1$ vertices, an arbitrary vertex of the graph is selected as start seed $S$. 

\item A $RangeSearch(\crEG_d, S, v, k_{\extend}, \varepsilon_{\extend})$ is performed for the new vertex $v$. 
The parameter $\varepsilon_{\extend}$ restricts the search range and $k_{\extend}$ is the size of the search result.

\item The best vertex $b$ of the search result not yet connected to the new vertex $v$ is chosen.

\item Based on a selection criterion depicted in Figure \ref{fig:add_vertex} and described later in this section, the edge $(b,n)$ is removed from $b$ if $n$ is not adjacent to $v$. 

\item By connecting $b$ and $n$ to the new vertex $v$, the connectivity and regularity of the graph is restored.

\item The steps 3-5 are repeated until $v$ has enough edges. 
\end{enumerate}

\noindent \textbf{Selection Scheme.} Figure \ref{fig:add_vertex} illustrates four ways to select vertex $n$ in step (4). 
Scheme (A) tries to connect the new vertex to the most similar other vertex. In (B) the shortest edge of the best vertex is removed, following the idea the incident vertex is likely to be similar and a good candidate for the new vertex. 
In (C), the longest edge of the best vertex is removed and the two incident vertices are connected to the new vertex. While half of the new neighbors may not be ideal for $v$, the quality of the existing neighborhoods is only slightly affected. 
Scheme (D) replaces the neighbor for which the resulting average neighbor distance of the graph is the lowest.

Although scheme (D) aims to improve the average neighbor distance at each step, its decisions impact the subsequent selection of vertex $b$ and therefore does not guarantee to always achieve the best overall results. Experiments in Section \ref{sec:evaluation} suggest scheme (C) is preferable for datasets with a high local intrinsic dimensionality, otherwise scheme (D) should be preferred. It should be noted the graph quality (GQ) for scheme (A), (C) and (D) remains the same in Figure \ref{fig:add_vertex}, making it unsuitable as a metric for small graph changes.

The complete graph extension procedure is described in Algorithm \ref{alg:extendGraph}. Since it is possible for all selected vertices $b$ and $n$ to be also elements of the search result, the minimum search result size $k_{\extend}$ should be at least $d$.
\\
\begin{algorithm}[h!]\small
    \caption{ExtendGraph($G, d, v, S, k_{\extend}, \varepsilon_{\extend}$)}
    \label{alg:extendGraph}
    \begin{algorithmic}[1]
    \Require current graph $G$, even degree $d \in \mathbb{N}$ and $d \geq 4$, new vertex $v$, set of seed vertices $S$, number of search results $k_{\extend} \in \mathbb{N}$ with $k_{\extend} \geq d$, search range factor $\varepsilon_{\extend} \in \mathbb{R}^+$
    \Ensure connect $v$ to the closest vertices in $G$
	\State $U \gets \emptyset$ \Comment{new neighbors of $v$}
	\State $S \gets$ RangeSearch($G, S, v, k_{\extend}, \varepsilon_{\extend}$)
	\Comment{find $k$ vertices similar to $v$}
    \State \textit{skipMRNG} $\gets$ false \Comment{two phases: with and w/o MRNG checks}
    \While {$|U| < d$} \Comment{stop if enough neighbors have been found}
        \State $B \gets S \setminus U$ \Comment{not yet connected candidates}
        \While {$|U| < d$ AND $B \neq \emptyset$} \Comment{check candidates}   
            \State $b \gets \argmin_{x \in B} \delta(x, v)$ \Comment{closest candidate to $v$ in $B$}
            \State $B \gets B \setminus \{b\}$ \Comment{ignore candidate next iteration}
            
            \If{\textit{skipMRNG} $=$ true OR checkMRNG($G, v, b$)} 
                \State $N \gets N(G,b) \setminus U$ \Comment{another candidate connected to $b$}
                \State $n \gets$ vertex in $N$ selected by scheme (A, B, C or D)
                \State $U \gets U \cup \{n\} \cup \{b\}$ \Comment{add to list of new neighbors of $v$}
                \State remove edge $(n,b)$
                \State add edge $(v,n)$ and $(v,b)$
            \EndIf
        \EndWhile
        \State \textit{skipMRNG} $\gets$ true \Comment{second phase without MRNG checks}
    \EndWhile
    \end{algorithmic}    
\end{algorithm}

\noindent \textbf{Approximating MRNG} The properties of the crEG and the described construction process make the EG an approximation of a \textit{Delaunay Graph} (DG) and a \textit{Relative Neighborhood Graph} \cite{Malkov2014}. Additionally Algorithm \ref{alg:checkMRNG} is used to identify potential neighbors which are part of a \textit{Monotonic Relative Neighborhood Graph} (MRNG). Depending on $k_{\extend}$ and the distribution of the data points, there may not be enough neighbor candidates satisfying the MRNG requirements. If $|U| < d$ after the first pass in Algorithm \ref{alg:extendGraph}, the MRNG tests are disabled and the retrieval of good $b$ vertices is repeated.

\begin{algorithm}\small
    \caption{checkMRNG($G, v1, v2$)} 
    \label{alg:checkMRNG}
    \begin{algorithmic}[1]
    \Require graph $G$, vertex $v1 \in V$ of $G$, vertex $v2 \in V$ of $G$
    \Ensure an edge between $v1$ and $v2$ is MRNG conform
    \ForAll {$u \in N(G, v1) \cap N(G, v2)$}
        \If{$\delta(v1, v2) > max(w_{v1, u}, w_{v2, u})$)} 
            \State \Return false
        \EndIf
    \EndFor
    \State \Return true
    \end{algorithmic}     
\end{algorithm}

\subsection{Continuous Edge Optimization} \label{sec:continuousEdgeOptimizations}

\begin{figure}[t!]
\centering
\includegraphics[trim=0.5cm 1.3cm 0cm 1.5cm,clip=true,width=\linewidth]{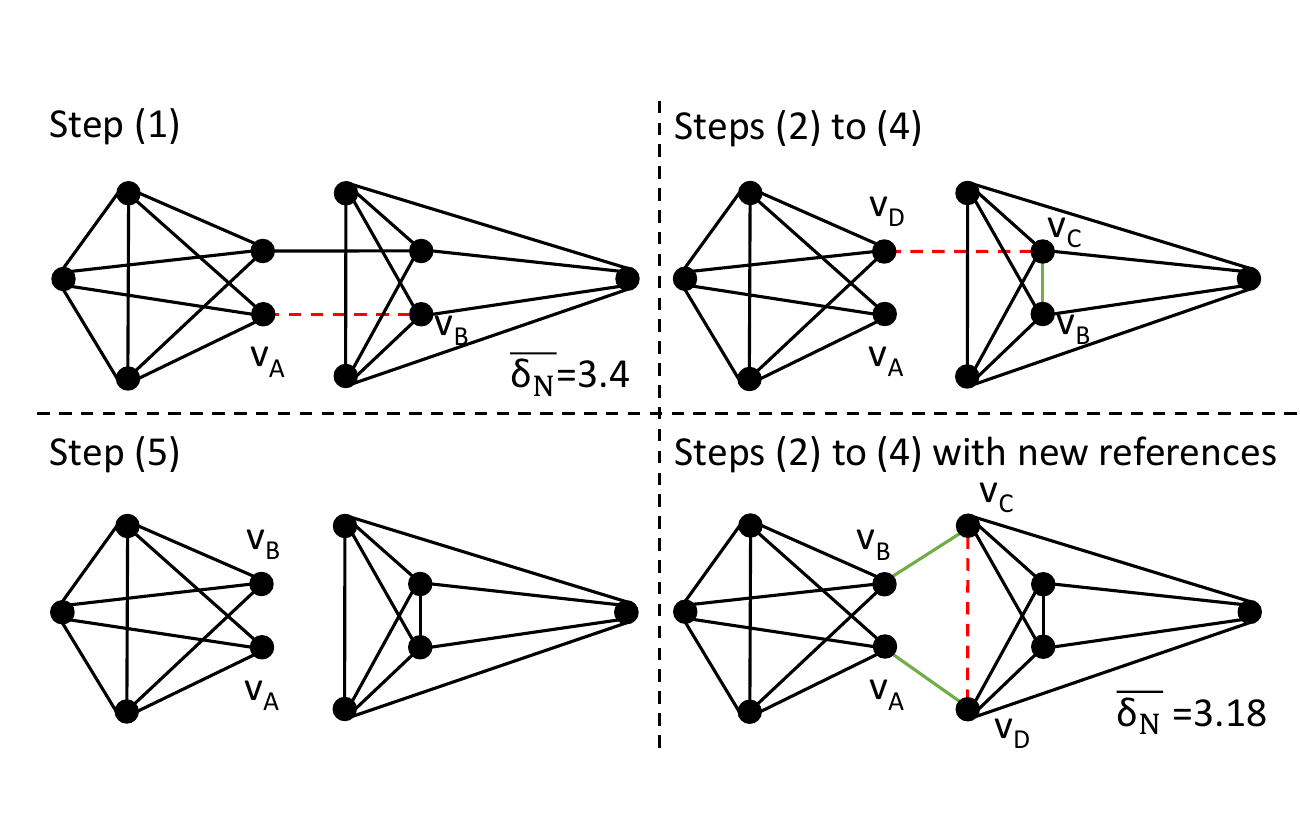}
\captionsetup{font=small}
\vspace{-0.5cm}
\caption{The continuous edge optimization process is illustrated by an 2D example. Starting from a valid $\crEG_4$ in Step(1) the algorithm tries to replace the edge $(\va, \vb)$ with a better edge constellation.} \label{fig:continuous_edge_optimization}

\end{figure}

During graph expansion, new vertices may replace existing short edges, causing some vertices to lose favorable connections. To mitigate this problem, we developed a \textit{continuous edge optimization} algorithm designed to minimize the \textit{average neighbor distance} in an existing crEG. The ability to repair suboptimal edges is crucial for future vertex removal strategies, which are not covered in this paper. Although multi-threading this algorithm is challenging, it can still operate in parallel with regular user search queries in a production system. The optimization process for an edge $(\va, \vb)$ is detailed below and illustrated in Figure \ref{fig:continuous_edge_optimization}:

\begin{enumerate}
    \item \textbf{Remove Edge:} Delete $(\va, \vb)$ which might result in the loss of the 2-edge connectivity. However, there is still one other path between $\va$ and $\vb$.
    
    \item \textbf{Find Replacement:} A RangeSearch with $S = \{\va\}$ is performed to find a suitable neighbor for $\vb$. From the result set, vertex $\vc$ and its neighbor $\vd$ are selected such that $\gain = \delta(\va, \vb) - \delta(\vb, \vc) + \delta(\vc, \vd)$ is maximized while ensuring $\vc \neq \va$, $\vc \neq \vb$, and $\vd \neq \vb$, $N(G, \vb) \cap {\vc} = \emptyset$.  

    \item \textbf{Edge Swap:} Replace $(\vc, \vd)$ with $(\vb, \vc)$. The vertices $\{\vb, \vc\}$ may become unreachable for $\va$ and $\vd$.
    
    \item \textbf{Restore:} Try to restore regularity and connectivity.
    \begin{description}  
         \item[Case a:] If $\va = \vd$, the vertex is missing two edges.  
         A RangeSearch with $S = \{\vb, \vc\}$ for query $\va$ is performed, $\ve$ is selected from the result set and $\vf$ from its neighborhood such that $\va \neq \ve$, $\va \neq \vf$, $N(G, \va) \cap \{\ve, \vf\} = \emptyset$ and $(\gain +\ \delta(\ve, \vf) - \delta(\va, \ve) - \delta(\va, \vf))$ is maximized. If the final gain is positive, the edge $(\ve, \vf)$ is replaced with the two edges $(\va, \ve)$ and $(\va, \vf)$.         
         \item[Case b:] If $\va \neq \vd$, the vertices $\va$ and $\vd$ can be connected if: $N(G, \va) \cap {\vd} = \emptyset$, $\gain -\ \delta(\va, \vd) > 0$ and there is a path from $\va$ or $\vd$ to $\vb$ or $\vc$.
    \end{description}
       
    \item \textbf{Recursion (if needed):} If $\va$ and $\vd$ are still missing an edge, repeat steps (2) to (4) by referencing $\vb$ to the vertex previously denoted by $\vd$ and starting the neighborhood search of step (2) at the two previous vertices $\vb$ and $\vc$. 
     
    \item \textbf{Termination:} If no solution is found after a few iterations (typically 5 are enough), all previous changes are reverted.
\end{enumerate}

\noindent Swapping the vertex labels and search seeds in step (5) ensures that step (3) will always reconnect the two potential graph components of the last iteration. Regardless of how the algorithm ends, the final graph retains its original properties. 
The entire process is detailed in Algorithm \ref{alg:optimizeEdge} and when paired with Algorithm \ref{alg:continuousEdgeOptimization}, it iteratively minimizes average neighbor distance. 
These changes can increase the search efficiency as shown in Section \ref{sec:qualityOfEdges}.

\begin{algorithm}[h]\small
    \caption{optimizeEdge($G, \va, \vb, i_{\opt}, k_{\opt}, \varepsilon_{\opt}$)} \label{alg:optimizeEdge}
    \begin{algorithmic}[1]
    
    \Require current graph $G$, edge between $\va$ and $\vb$, max number of changes $i_{\opt}$, search parameters $k_{\opt}$ and $\varepsilon_{\opt}$
    \Ensure try to improve the edge $(\va, \vb)$ 
    \State $M \gets \emptyset$ \Comment{history of modifications}
    \State $\gain \gets w_{\va, \vb}$ \Comment{change would improve $\AND$}
    \State remove edge $(\va, \vb)$ from $G$ and add to $M$ \Comment{Step 1}
    \State $\vc \gets \va, \vd \gets \va$ \Comment{needed for multiple iterations}
    \While {$|M| < i_{\opt}$}

        \State $b \gets \gain$ \Comment{best current gain}
        \State $S \gets$ RangeSearch($G, \{\vc, \vd\}, \vb, k_{\opt}, \varepsilon_{\opt}$) \Comment{Step 2}
        \ForAll {$s \in S$} \Comment{find a new neighbor s for $\vb$}
            \If {$s \neq \va$ and $s \neq \vb$ and $N(G, \vb) \cap \{s\} = \varnothing$}
                \ForAll {$n \in N(G, s)$} \Comment{select a bad edge of s}
                    \If {$n \neq \vb$ and $b < \gain - \delta(s, \vb) + w_{s, n}$}
                        \State $b \gets \gain - \delta(s, \vb) + w_{s, n}$
                        \State $\vc \gets s$, $\vd \gets n$
                    \EndIf
                \EndFor
            \EndIf
        \EndFor
        \If {$b == \gain$} \Comment{all potential neighbors in $S$ do not improve $\AND$}
            \Break
        \EndIf        
        \State $\gain \gets b$
        \State add edge $(\vb, \vc)$ to $G$ and add to $M$ \Comment{Step 3: swap edges}
        \State remove edge $(\vc, \vd)$ from $G$ and add to $M$            
        
        \If {$\va == \vd$} \Comment{Step 4a: $\va$ missing two edges}
            \State $b \gets 0$ \Comment{best current gain}
            \State $S \gets$ RangeSearch($G, \{\vb, \vc\}, \va, k_{\opt}, \varepsilon_{\opt}$) 
            \ForAll {$s \in S$} \Comment{find good neighbor for $\va$}
                \If {$s \neq \va$ and $N(G, \va) \cap \{s\} = \varnothing$}
                    \ForAll {$n \in N(G, s) \setminus \{\va\}$}
                        \If {$b < \gain + w_{s, n} - \delta(s, \va) - \delta(n, \va)$}
                            \State $b \gets \gain + w_{s, n} - \delta(s, \va) - \delta(n, \va)$
                            \State $\ve \gets s$, $\vf \gets n$
                        \EndIf
                    \EndFor
                \EndIf
            \EndFor  
            \If {$b > 0$} \Comment{place $\va$ in between two adjacent vertices}
                \State $\gain \gets b$
                \State remove edge $(\ve, \vf)$ from $G$
                \State add edge $(\va, \ve)$ and $(\va, \vf)$ to $G$
                \State \Return 
            \EndIf   

        \ElsIf {$N(G, \va) \cap \{\vd\} = \varnothing$ and $0 < (gain - \delta(\va, \vd))$}
            \State $S_A \gets$ RangeSearch($G, \{\vb, \vc\}, \va, k_{\opt}, \varepsilon_{\opt}$) \Comment{Step 4b}
            \State $S_D \gets$ RangeSearch($G, \{\vb, \vc\}, \vd, k_{\opt}, \varepsilon_{\opt}$) \Comment{find path}
            \If {$S_A \cap \{\va\} = \{\va\}$ or $S_D \cap \{\vd\} = \{\vd\}$}    
                \State add edge $(\va, \vd)$ to $G$ \Comment{found a path, add edge}
                \State \Return 
            \EndIf
        \EndIf

        \State $v \gets \vd$, $\vd \gets \vc$, $\vc \gets \vb$, $\vb \gets v$ \Comment{Step 5: prepare next iter.}
    \EndWhile
    \State revert changes of G listed in $M$
    \end{algorithmic} 
\end{algorithm}
\vspace{-0.5cm}
\begin{algorithm}[h]\small
    \caption{continuousEdgeOptimization($G, i_{\opt}, k_{\opt}, \varepsilon_{\opt}$)} \label{alg:continuousEdgeOptimization}
    \begin{algorithmic}[1]
    	\Require graph $G$, max no. of changes $i_{\opt}$, search parameters $k_{\opt}$ \& $\varepsilon_{\opt}$
        \Ensure improve the average neighbor distance by swapping edges
        
        \State $v1 \gets rnd(G)$ \Comment{random vertex in $G$}
        \State $N \gets N(G, v1)$ \Comment{neighbors of $v1$}

        \ForAll {$v2 \in N$}
            \If {$checkMRNG(G, v1, v2) == False$}
                \State $optimizeEdge(G, v1, v2, i_{\opt}, k_{\opt}, \varepsilon_{\opt})$  
            \EndIf   
        \EndFor  
    \end{algorithmic} 
\end{algorithm}

\section{Evaluation}
\label{sec:evaluation}

In order to ensure a comprehensive comparison, we evaluated our implementation of \textbf{crEG} and a version without edge optimization (\textbf{EG}) against several state-of-the-art graph-based algorithms: \textbf{kGraph} \cite{GithubKGraph}, \textbf{EFANNA} \cite{GithubEFANNA}, \textbf{DPG} \cite{GithubDPG}, \textbf{ONNG} \cite{GithubNGT}, \textbf{HNSW} \cite{GithubHNSW}, \textbf{NSG} \cite{GithubNSG}, \textbf{NSSG} \cite{GithubSSG} in regards to their index build time (including all auxiliary data structures), trade-off between search speed and recall rate, as well as peak memory usage during construction and search. The serial scan results from FAISS \cite{Johnson2019} serves as a baseline.

All experiments were conducted on a Ryzen 2700x @ 4GHz and 64GB of DDR4 memory @ 2133MHz. Additionally, the graph indices were maintained in memory to avoid any I/O bottlenecks. The multimedia datasets used for the evaluation are frequently cited in related literature \cite{Wang2021Survey, Cong2021, DPG2020, Chen2023} and are listed in Table \ref{tab:datasets}. The local intrinsic dimension (LID) \cite{Facco2018} for each dataset is computed to better reflect the difficulty of its data distribution.

\begin{table}[h]
    \centering
    \begin{tabularx}{\linewidth} { 
   >{\arraybackslash}X 
   >{\arraybackslash}r  
   >{\arraybackslash}r 
   >{\arraybackslash}r 
   >{\arraybackslash}r  }
        \hline
        \textbf{Dataset} & $\bm{m}$ & \mbox{\textbf{\# Base}} & \mbox{\textbf{\# Query}} & \textbf{LID}\\ 
        \hline
        Audio \cite{Audio} & 192 & 53,387 & 200 & 14.2\\
        SIFT1M \cite{Jegou2011} & 128 & 1,000,000 & 10,000 & 9.2\\
        Deep1M \cite{Babenko2016} & 96 & 1,000,000 & 10,000 & 15.5\\       
        GloVe \cite{Pennington2014} & 100 & 1,183,514 & 10,000 & 21.7\\ 
        \hline
    \end{tabularx} 
    \vspace{0.5em}
    \caption{Details of the used datasets: dimensionality of the feature vectors (m), number of data points (\# Base) and queries (\# Query), and local intrinsic dimension of the dataset (LID).}
    \label{tab:datasets}
    \vspace{-2.0em}
\end{table}

\subsection{Comparing the search performance} \label{sec:searchExperiments}

\begin{figure*}[ht!]
    \begin{subfigure}{0.24\textwidth}
        \includegraphics[trim=2.0cm 4.0cm 2.0cm 6.5cm,clip=true,width=\textwidth]{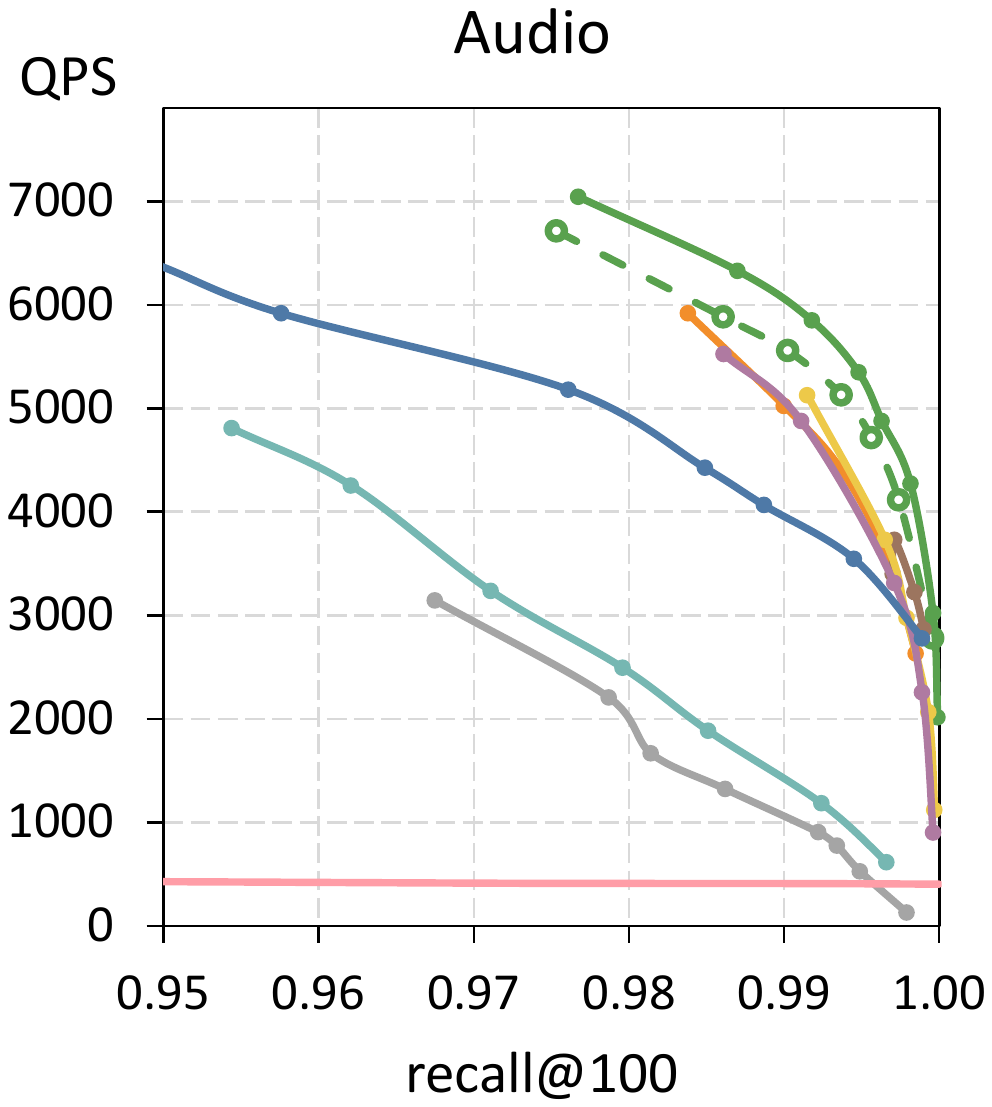}
    \end{subfigure}
    \hfill
    \begin{subfigure}{0.24\textwidth}
        \includegraphics[trim=2.0cm 4.0cm 2.0cm 6.5cm,clip=true,width=\textwidth]{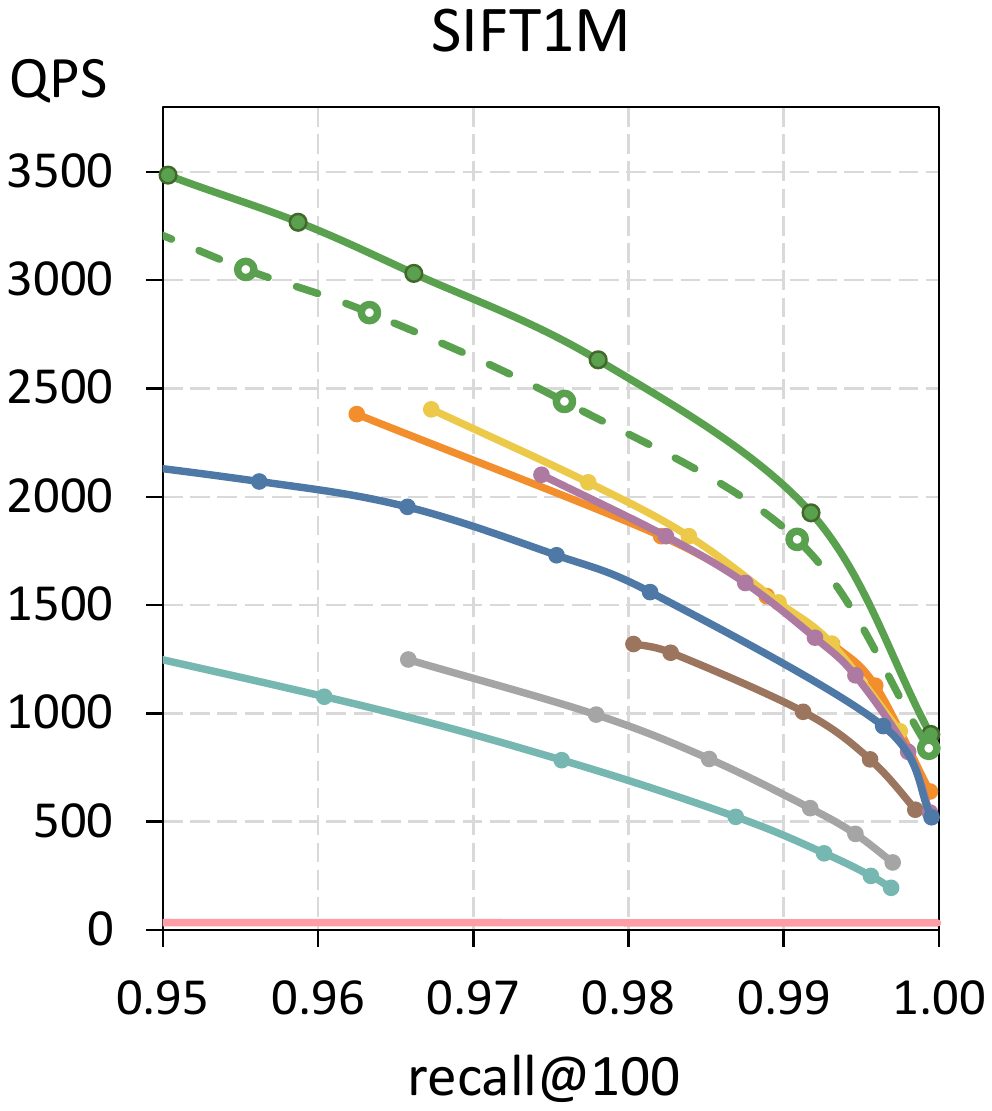}
    \end{subfigure}
    \hfill
    \begin{subfigure}{0.24\textwidth}
        \includegraphics[trim=2.0cm 4.0cm 2.0cm 6.5cm,clip=true,width=\textwidth]{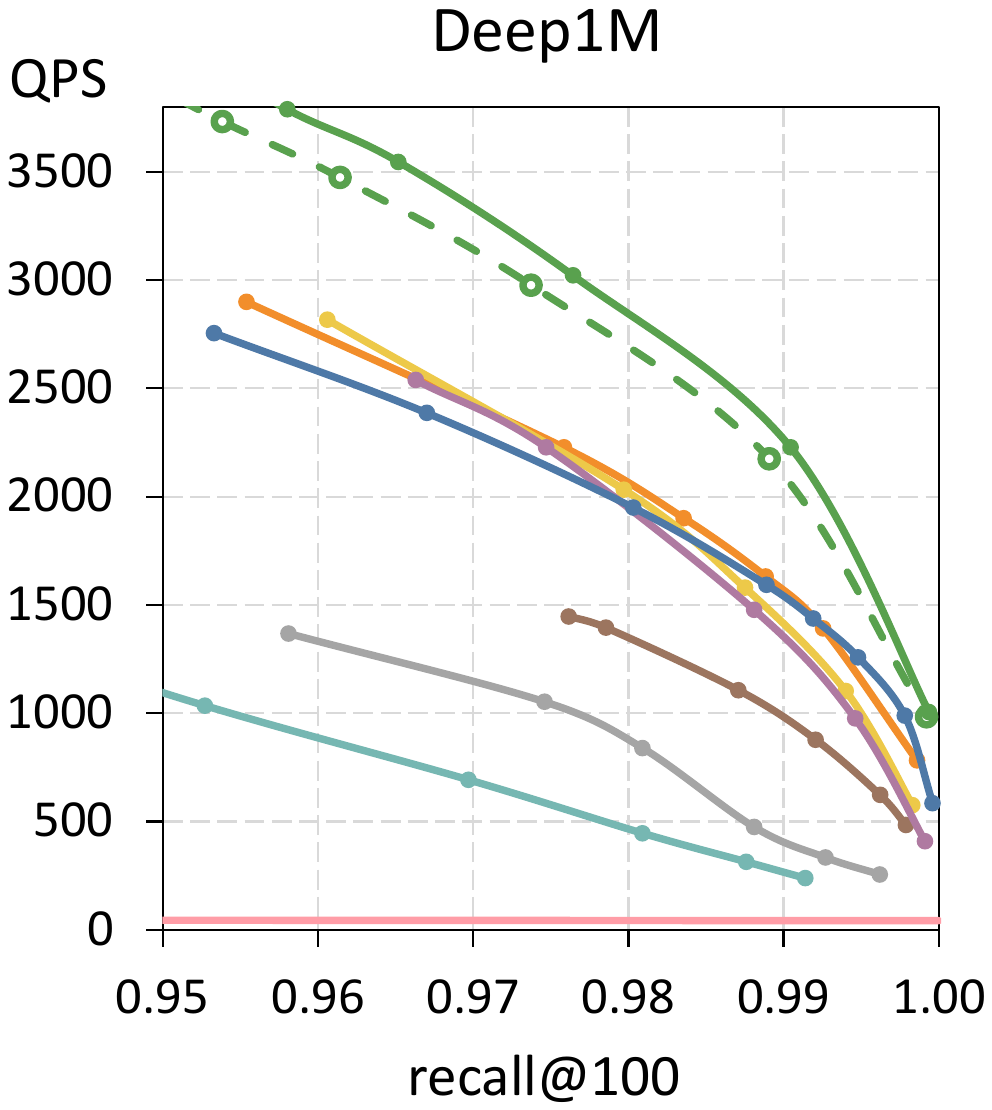}
    \end{subfigure}
    \hfill
    \begin{subfigure}{0.24\textwidth}
        \includegraphics[trim=2.0cm 4.0cm 2.0cm 6.5cm,clip=true,width=\textwidth]{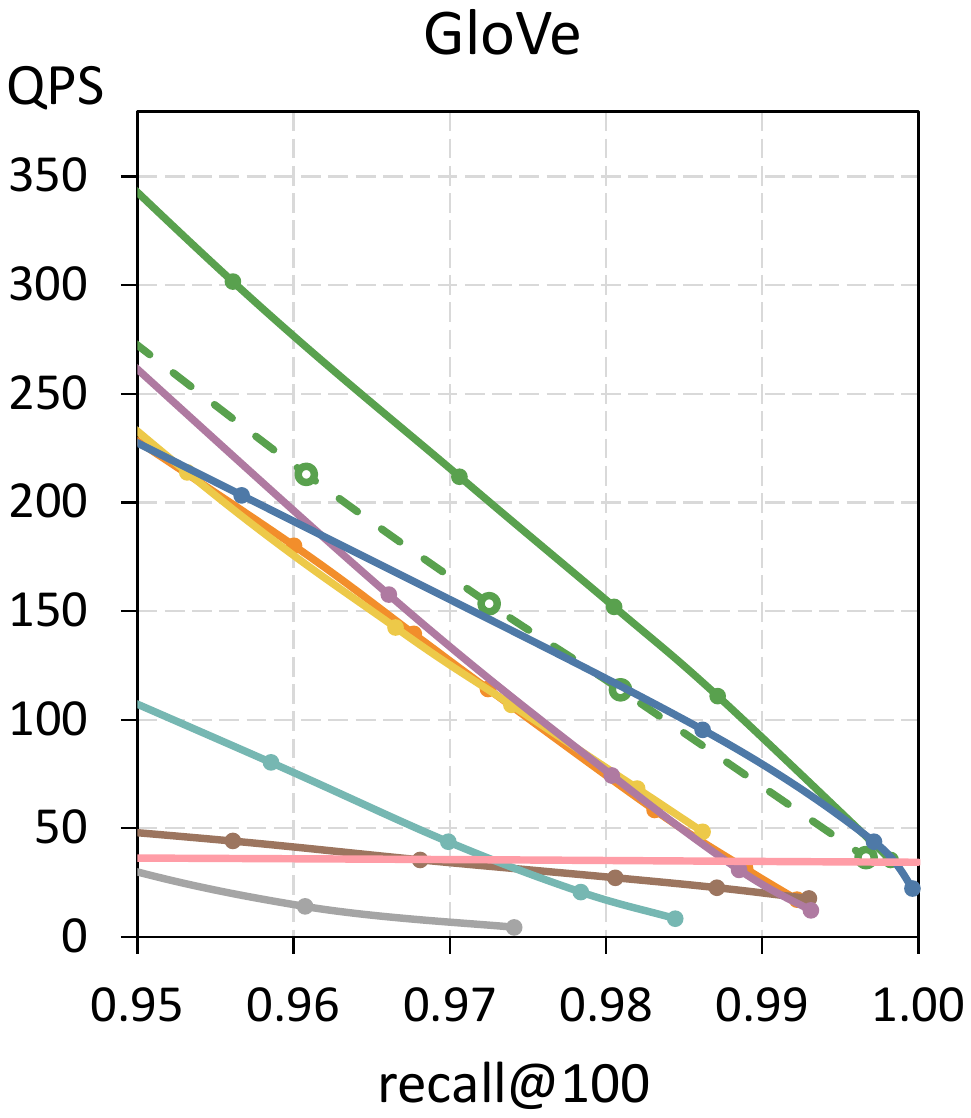}
    \end{subfigure}
    \begin{subfigure}{0.95\textwidth}
        \includegraphics[trim=1.75cm 13.80cm 2cm 15.6cm,clip=true,width=\textwidth]{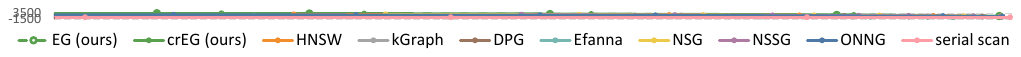}
    \end{subfigure}
    \vspace{-0.2cm}
    \caption{The number of queries per second (QPS) in relation to recall@100 for approximate nearest neighbor search was tested for different graphs and datasets (top right is better). The proposed \textit{continuously refining Exploration Graph} (crEG) and its non-optimized version \textit{Exploration Graph} (EG) is more efficient than the current state of the art. }
    \label{fig:search_performance}
\end{figure*}

\begin{table*}[ht!]
    \centering
    \begin{tabularx}{\textwidth} { |
   >{\centering\arraybackslash}l  |
   >{\centering\arraybackslash}X  
   >{\centering\arraybackslash}X   
   >{\centering\arraybackslash}X  
   >{\centering\arraybackslash}X  |
   >{\centering\arraybackslash}X  
   >{\centering\arraybackslash}X   
   >{\centering\arraybackslash}X  
   >{\centering\arraybackslash}X  |
   >{\centering\arraybackslash}X  
   >{\centering\arraybackslash}X   
   >{\centering\arraybackslash}X  
   >{\centering\arraybackslash}X  |
   >{\centering\arraybackslash}X  
   >{\centering\arraybackslash}X   
   >{\centering\arraybackslash}X  
   >{\centering\arraybackslash}X  |}
        \hline
        \multirow{2}{*}{

        \parbox{1.7cm}
        { \centering
        \ \\[-0.7ex]
        \textbf{Dataset}
        \ \\[2.6ex]
        \textbf{Algorithm}
        }
        
        } &
        \multicolumn{4}{c|}{\textbf{Audio (41 MB)}} &
        \multicolumn{4}{c|}{\textbf{SIFT1M (516 MB)}} &
        \multicolumn{4}{c|}{\textbf{Deep1M (388 MB)}} &   
        \multicolumn{4}{c|}{\textbf{GloVe (478 MB)}} \\ 
        \cline{2-17}
        &
        $\bm{IT}$ \textbf{(min)} & 
        $\bm{PM_c}$ \textbf{(MB)} & 
        $\bm{PM_s}$ \textbf{(MB)} & 
        $\bm{FS}$ \textbf{(MB)} &
        $\bm{IT}$ \textbf{(min)} & 
        $\bm{PM_c}$ \textbf{(MB)} & 
        $\bm{PM_s}$ \textbf{(MB)} & 
        $\bm{FS}$ \textbf{(MB)} &
        $\bm{IT}$ \textbf{(min)} & 
        $\bm{PM_c}$ \textbf{(MB)} & 
        $\bm{PM_s}$ \textbf{(MB)} & 
        $\bm{FS}$ \textbf{(MB)} &
        $\bm{IT}$ \textbf{(min)} & 
        $\bm{PM_c}$ \textbf{(MB)} & 
        $\bm{PM_s}$ \textbf{(MB)} & 
        $\bm{FS}$ \textbf{(MB)} \\ 
        \hline
        EG (ours)& \textbf{0.2} & \textbf{52} & \textbf{48} & 50 & \textbf{6.3} & \textbf{780} & \textbf{665} & 756 & \textbf{5.6} & \textbf{653} & \textbf{540}  & 628 & \textbf{16.5} & \textbf{821} & 678 & 762 \\
        crEG (ours)& 0.5 & \textbf{52} & \textbf{48} & 50 & 17.4 & \textbf{780} & \textbf{665} & 756 & 21.5 & \textbf{653} & \textbf{540} & 628 & 46.7 & \textbf{821} & 678 & 762 \\
        HNSW & 0.7 & 67 & 67 & 53 & 35.8 & 892 & 892 & 660 & 34.9 & 766 & 806 & 604 & 54.2 & 985 & 985 & 781 \\   
        kGraph & \textbf{0.2} & 120 & 86 & 79 & 15 & 3984 & 1656 & 1568 & 16.7 & 4324 & 1656 & 1568 & 41.8 & 6111 & 2674 & 2139 \\
        DPG & 0.7 & 203 & 69 & 48 & 17.2 & 4085 & 1116 & 707 & 16.8 & 4188 & 1378 & 581 & 26.5 & 4753 & 1886 & 926 \\
        EFANNA & 0.6 & 240 & 55 & 50 & 10.6 & 2356 & 748 & 720 & 10.2 & 2206 & 620 & 592 & 83.7 & 8410 & 2447 & 2376 \\
        NSG & 1.3 & 227 & 50 & \textbf{45} & 30.6 & 3797 & 671 & \textbf{635} & 29.6 & 3658 & 554 & \textbf{520} & 112.1 & 8410 & \textbf{603} & \textbf{544} \\
        NSSG & 1.9 & 380 & 50 & \textbf{45} & 22.4 & 4017 & 710 & 686 & 21.4 & 3864 & 583 & 548 & 90.5 & 8410 & 642 & 583 \\
        ONNG & 4.2 & 297 & 116 & 60 & 292.4 & 7297 & 1112 & 246.9 & 9513 & 1948 & 830 & 947 & 3644.5 & 6588 & 1612 & 1282 \\
        \hline
    \end{tabularx}
    \vspace{0.1cm}
    \caption{Single threaded indexing time ($\bm{IT}$) and peak memory consumption including the feature vectors during construction ($\bm{PM_c}$) and search ($\bm{PM_s}$) of graph-based approaches together with their resulting file size ($\bm{FS}$)}
    \label{tab:indexing_speed_and_memory}
    \vspace{-0.5cm}
\end{table*}

The datasets are organized in base and query data. The assessed systems index the feature vectors of the base data and subsequently retrieve the $k$ most similar feature vectors from the index for all query data. The average $\recallAtVar{k}$ rate and the search speed in queries per seconds (QPS) are determined for different parameters of the search algorithms. We focused on achieving a high search efficiency for recall rates above 95\% and modified the parameters accordingly. The graph construction parameters were taken from the original papers or this survey \cite{Wang2021Survey} and are documented on our project page\footnote{\repourl}. The FAISS (serial scan) curve was created with a reduced base data set to obtain correct times for lower recall rates.
\\
\\
\textbf{Observations:}
Our analysis reveals a close alignment between the results in Figure \ref{fig:search_performance} and the findings reported in the literature \cite{Cong2021, Wang2021Survey}. When considering a recall rate of 99\%, the new \textit{continuously refining Exploration Graph} outperforms current state-of-the-art algorithms HNSW, SSG, and NSG in terms of search speed. Specifically, it processes queries 20\% faster for the Audio, 30\% for the SIFT1M, 50\% for the Deep1M, and 250\% for the GloVe dataset. The performance gap widens for datasets with a high local intrinsic dimensionality. The Audio dataset is an outlier due to its smaller size, resulting in less pronounced differences. Remarkably, even the \textit{Exploration Graph} (EG), without any edge optimization, is very competitive and trailing only slightly behind the ONNG for the GloVe dataset, which required over 200 times longer to build.

\subsection{Run-time and memory consumption} \label{sec:constructionTimeAndMemoryConsumption}
When designing the \textit{Exploration Graph}, our primary goal was to create a graph with high search efficiency that can be constructed quickly. A further condition was to be competitive in terms of memory consumption. Table \ref{tab:indexing_speed_and_memory} documents the single-threaded indexing time (IT) and peak memory (PM) requirements of all tested graphs. The later was measured during the construction phase ($PM_C$) and the subsequent test phase ($PM_S$). All numbers include the size of the feature vectors and should give a general idea of the memory requirements of the different approaches. 

After storing the graph to disk, the file size (FS) is measured, which again includes the feature vectors from the base dataset. While aiming to use minimal disk space, it is important to consider the size of the dataset in relation to the file size of the graph. For feature vectors with thousands of dimensions, a graph with 2-3 dozen edges per vertex only accounts for a small fraction of the total memory requirement.
\\
\\
\textbf{Observations:}
During the construction phase, memory usage tends to be higher for refinement graphs like DPG, NSG, NSSG, and ONNG, as they keep the initial KNNG and its pruned version in memory. In contrast, the crEG takes an incremental approach and has the lowest peak memory utilization during construction ($PM_C$). Its index file contains edge weights for future edge optimization steps, but omits them during the search phase.

Incremental graphs (e.g. HNSW and crEG) and k-NN graphs (e.g. kGraph and EFANNA) generally exhibit lower indexing times (IT) than graphs which prune edges. Although the pruning process is often quite fast, it necessitates an initial graph with numerous edges, which may take a significant amount of time to construct. It should be noted, that while graphs like kGraph, EFANNA and DPG are among the fastest to build, they have also the slowest search performance. Our new \textit{Exploration Graph} without additional edge optimization is 2-3 times faster to construct than the current SOTA search graphs (HNSW, SSG, NSG) while providing better search efficiency.

\begin{figure*}[ht!]
    \begin{subfigure}{0.24\textwidth}
        \includegraphics[trim=2.0cm 4.0cm 2.0cm 6.5cm,clip=true,width=\textwidth]{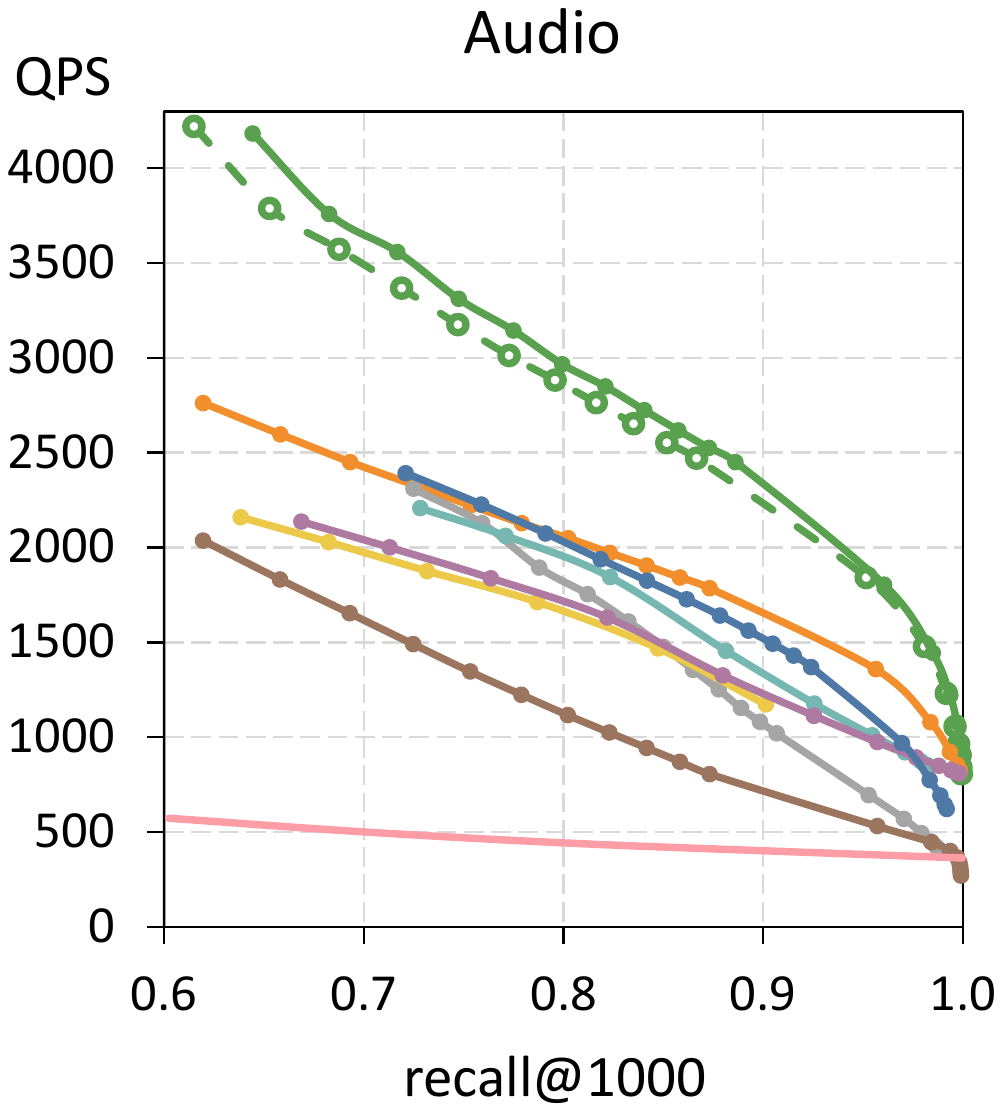}
    \end{subfigure}
    \hfill
    \begin{subfigure}{0.24\textwidth}
        \includegraphics[trim=2.0cm 4.0cm 2.0cm 6.5cm,clip=true,width=\textwidth]{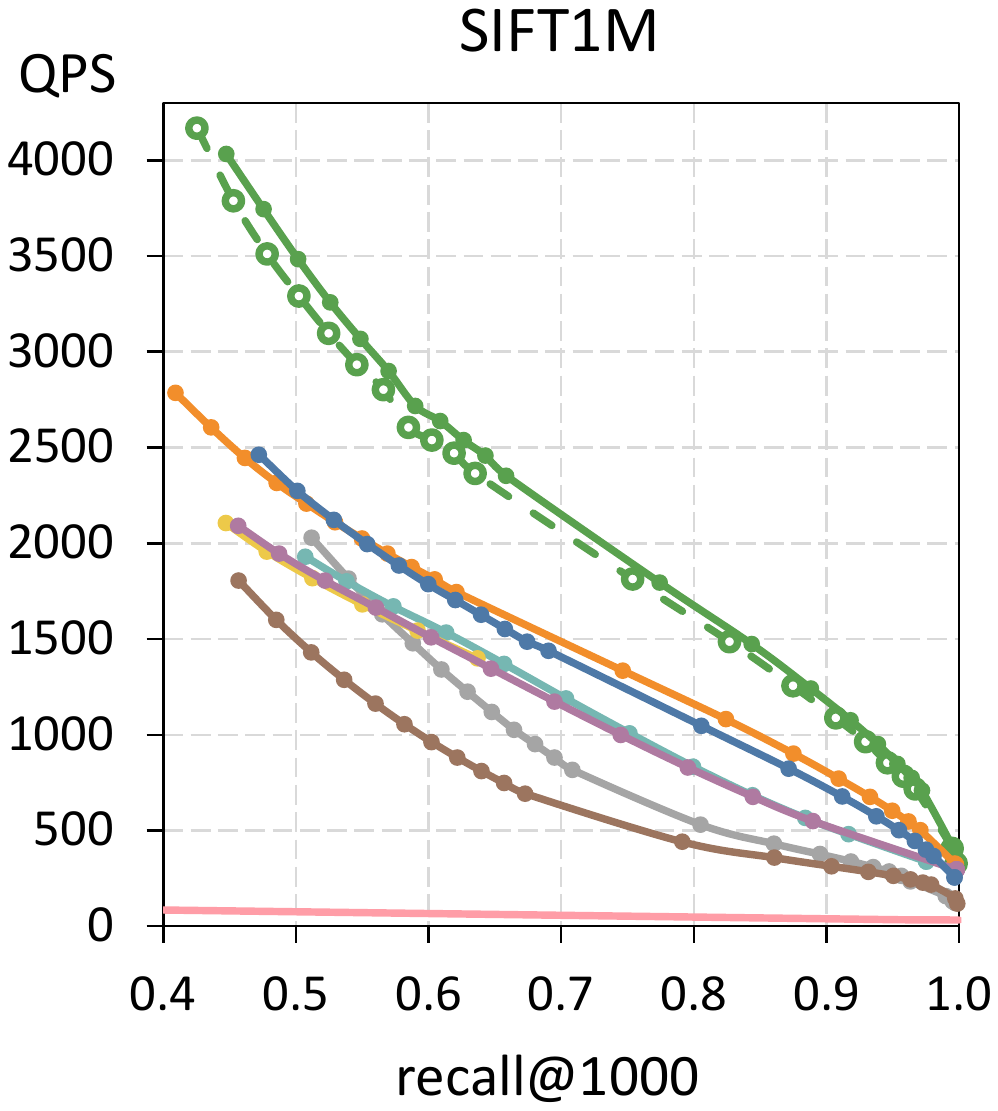}
    \end{subfigure}
    \hfill
    \begin{subfigure}{0.24\textwidth}
        \includegraphics[trim=2.0cm 4.0cm 2.0cm 6.5cm,clip=true,width=\textwidth]{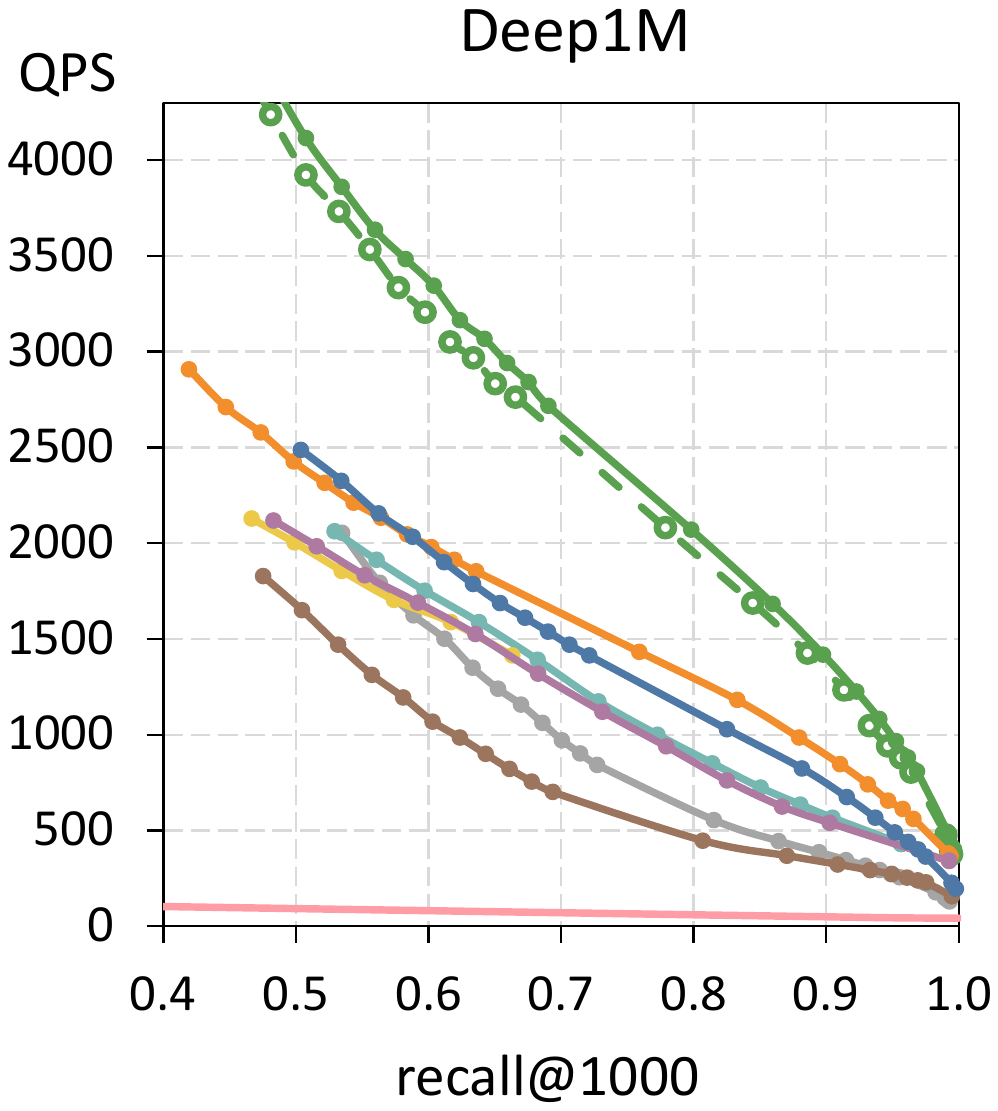}
    \end{subfigure}
    \hfill
    \begin{subfigure}{0.24\textwidth}
        \includegraphics[trim=2.0cm 4.0cm 2.0cm 6.5cm,clip=true,width=\textwidth]{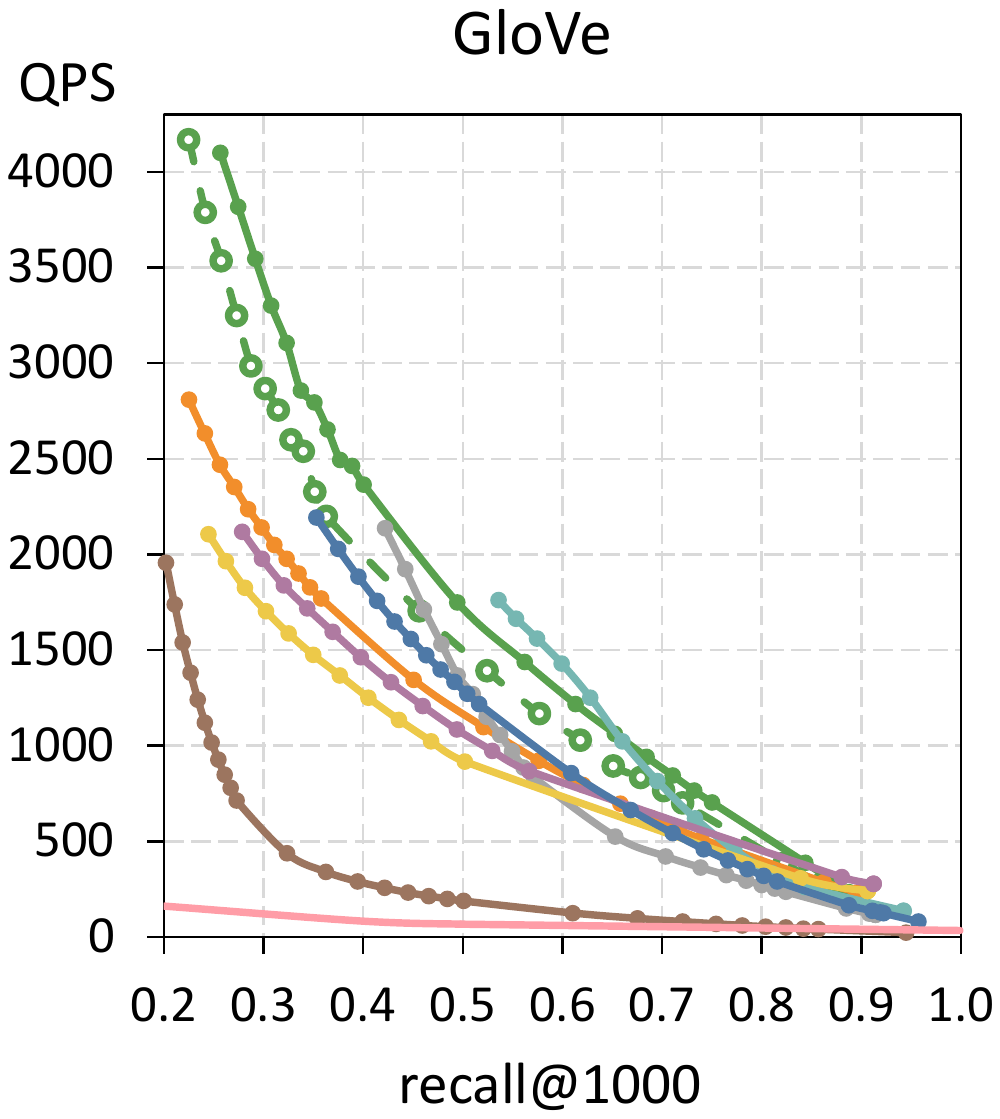}
    \end{subfigure}  
    \begin{subfigure}{0.95\textwidth}
        \includegraphics[trim=1.75cm 13.80cm 2cm 15.6cm,clip=true,width=\textwidth]{figures/recall_vs_qps_legend.pdf}
    \end{subfigure}
    \vspace{-0.2cm}
    \caption{Exploration performance measured by recall@1000 in relation to the number of queries per second (QPS). Various graphs are compared on four datasets with different hyper-parameter settings. Top right is better.}
    \label{fig:explore_performance}
    
\end{figure*}

\subsection{Exploration Quality} \label{sec:explorationExperiements}

The experiments described in Section \ref{sec:searchExperiments} adhere to the established testing protocols for approximate k-NN search, as outlined in \cite{Aumuller2020,Wang2021Survey,DPG2020}. These focus on identifying data points within the graph that are similar to a non-indexed query, yet they overlook another type of search that is particularly effective for graphs.

In the following experiments, each query represents a vertex in the graph. Additionally, the seed vertex for every graph search is set to the query's corresponding vertex. This search methodology is crucial in systems like recommendation engines, which need to find many similar items for a specific item, or in user-driven navigation tools where search inputs are iteratively refined through a feedback loop. In these scenarios, it is important to ensure that users are not shown items they have already encountered. Therefore, a comprehensive list of search results is often necessary. We refer to such searches as \textit{exploratory search}.

The graphs and search algorithms tested in this section are identical to those in Section \ref{sec:searchExperiments}. Only the queries and the seed vertices have been changed. For each dataset 10,000 vertices from the base dataset where randomly chosen to form the respective test set of the exploration tasks. The 1,000 most similar neighbors are searched per query and the entire recall range is considered. 

Ideally, $k$ is higher than the average number of edges per vertex of any tested k-NN graphs, as otherwise the neighborhood of the seed vertex is sometimes enough to achieve a recall rate of 100\%, which leads to false expectations for longer result lists. Additionally, the number of edges per vertex and the hyperparameters determine the range that the Efficiency-Recall curve covers in the diagram. The reason some curves start later in Figure 2 is due to this circumstance.
\\
\\
\textbf{Observations:}
As depicted in Figure \ref{fig:explore_performance}, it is evident the crEG outperforms the other graphs consistently across various recall rates. Depending on the desired recall up to 50\% more queries per second compared to the second-best graph can be archived. In some low-recall ranges k-NN graphs like EFANNA and k-graph archive a slightly better performance due to their hundreds of edges per vertex and high \textit{graph quality} metric value. The advantage diminishes in the high-recall range due to the lack of effective navigation edges and the existence of numerous hub vertices. Furthermore, they exhibit several source vertices with zero incoming edges, which limits general reachability and require a lot of memory.

The ranking of the best performing graphs varies between exploration tasks and standard search tasks, indicating that the effectiveness of a graph in approximate nearest neighbor search does not directly translate to its exploration capabilities. The crEG is designed to perform efficiently in both scenarios. In an additional experiment involving randomly selected seed vertices, the search efficiency of the algorithms was more akin to the Approximate Nearest Neighbor Search (ANNS) experiments described in Section \ref{sec:searchExperiments} than to the exploration results. Furthermore, exploration tests with $k = 100$ yielded significantly faster retrieval results compared to general ANNS under the same settings.

\section{Ablation Study} \label{sec:ablation}

\subsection{Empirical Scalability} \label{sec:scalabilityAndComplexity}

\begin{figure}[h!]
    \begin{subfigure}{0.48\columnwidth}
        \includegraphics[trim=5cm 2cm 5.4cm 2.2cm,clip=true,width=\columnwidth]{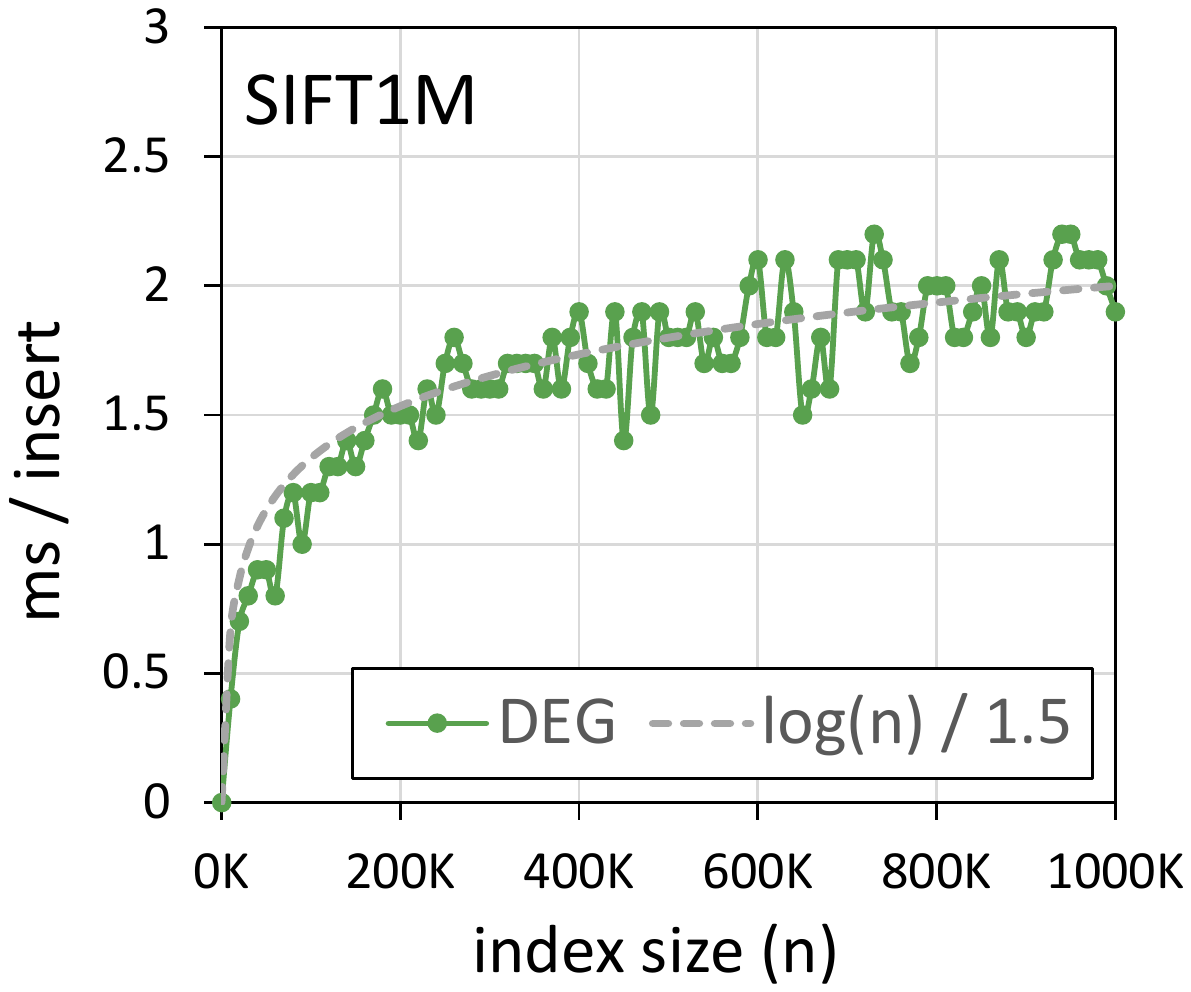}
    \end{subfigure}
    \hfill
    \begin{subfigure}{0.48\columnwidth}
        \includegraphics[trim=5cm 2cm 5.5cm 2.2cm,clip=true,width=\columnwidth]{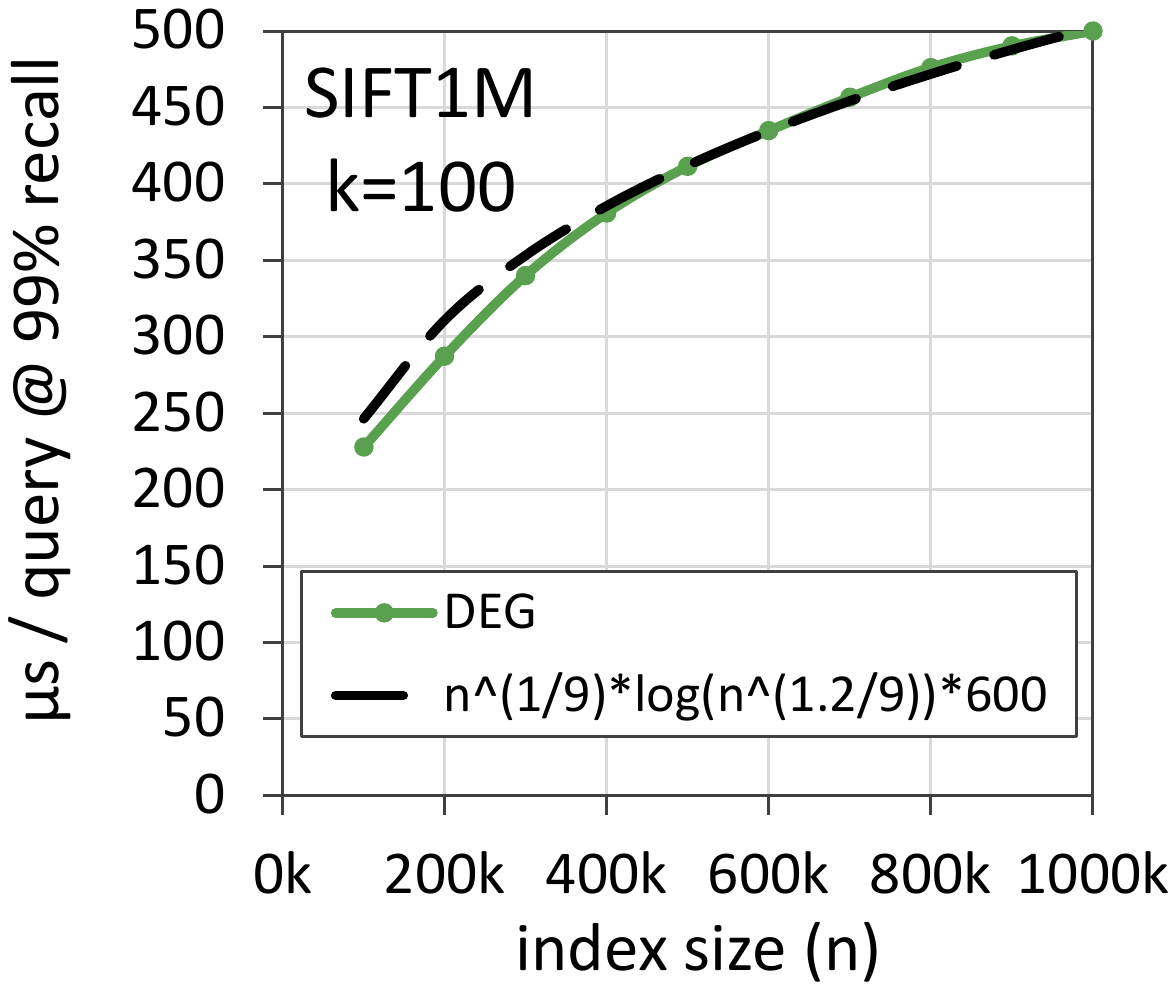}
    \end{subfigure}    
    
    \vspace{-1.0em}
    \caption{Influence of index size on indexing speed (left) and search speed (right) to achieve a 99\% recall rate at k=100.}   \label{fig:build_and_search_time_scaling}
\end{figure}

In our experiments, we found the optimal parameters of crEG do not change with the amount of data, as long as the data distribution stays the same. Therefore, we randomly draw samples of different sizes and create a crEG for each of these samples with the same parameters. 
The graph construction and search time to reach a recall of 99\% with $k = 100$ for all these subsets are recorded. 
The curves of Figure \ref{fig:build_and_search_time_scaling} show how the time increases with the number of vertices $n$ in the index for the SIFT1M dataset. 

The search time complexity concerning the number of vertices can be expressed as $O(n^\frac{1}{9} log(n^\frac{1}{9}))$. This is in line with the current state-of-the-art approaches HNSW, NSG and NSSG in \cite{Wang2021Survey}. 
The time complexity for adding a new vertex to the index is very similar to the search time complexity, as depicted in Figure \ref{fig:build_and_search_time_scaling} on the left. 
If the SIFT dataset had 1 billion data points, the time to add another vertex on the same hardware would be only about 6.6 ms.
The complexity for the other datasets is similar just with different constants.

\vfill\null

\subsection{Quality of edges} 
\label{sec:qualityOfEdges}
\begin{figure}[t!]
    \begin{subfigure}{0.49\columnwidth}
        \includegraphics[trim=0cm 8.2cm 23cm 0cm,clip=true,width=\columnwidth]{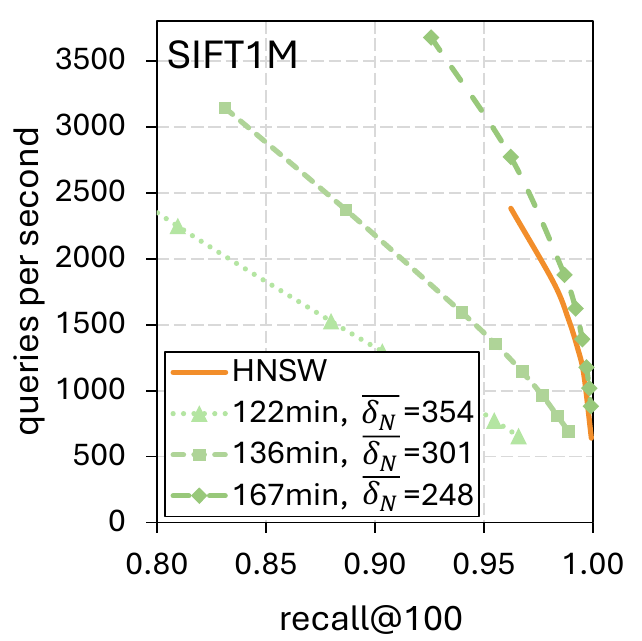}      
    \end{subfigure}
    \hfill
    \begin{subfigure}{0.49\columnwidth}
        \includegraphics[trim=2cm 5.5cm 2cm 7cm,clip=true,width=\columnwidth]{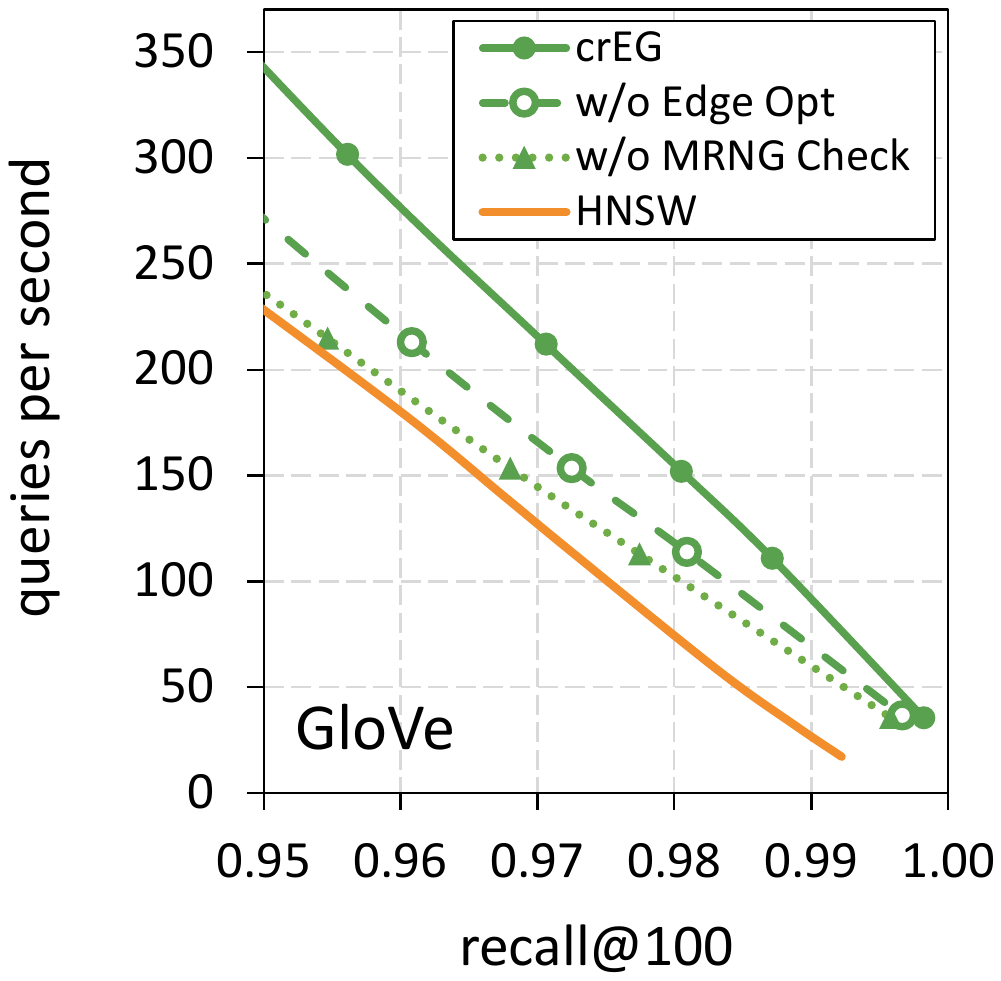}      
    \end{subfigure} 
    \vspace{-0.5em}
    \caption{Left: Edge optimization transforms a random even-regular undirected graph into an efficient crEG. $\AND$ represents the Average Neighbor Distance. Right: Influence of the different algorithmic components. HNSW (in orange) was added for reference.}  
    \label{fig:swap_scaling_sift}
\end{figure}

In order to investigates the effectiveness of edge optimization, an even-regular undirected graph with random edges was generated for the SIFT1M dataset and then optimized using Algorithm \ref{alg:continuousEdgeOptimization}.
As the algorithm improves the edges of random vertices, the graph's Average Neighbor Distance decreases and the search quality improves. 
The process was run single threaded and the search quality was evaluated after different numbers of iterations.
The results are shown in Figure \ref{fig:swap_scaling_sift}, where each curve represents a random graph optimized for a given duration. 
It took over two hours to get useful connections and then, after another half hour, the graph returned better search results than the state of the art. Although further edge optimization is possible, the return is diminishing.
Figure \ref{fig:swap_scaling_sift} on the right demonstrates the efficiency of the various algorithmic components discussed in this paper. Constructing the crEG without edge optimization results in just the EG, sacrificing search performance for faster construction time. On the other hand, forming the \textit{Exploration Graph} without MRNG checks decreases the search efficiency and only offers negligible time savings.

\section{Conclusion}
This paper introduces the \textit{continuously refining Exploration Graph} (crEG), an even-regular undirected graph manipulated by two algorithms: The first enables incremental graph expansion with new vertices in a quick manner. The second updates the neighborhood of existing vertices to reduce the \textit{average neighbor distance} and improve the overall search efficiency. 
The even regularity property and the undirected edges create a balance between short and long edges, essential for efficient search and exploration. 
Additionally, the crEG approximates MRNG, providing near logarithmic search and construction complexity. Extensive experiments have demonstrated the construction speed of EG is about 2-3 times faster than similar efficient graph systems. After refining the edges, crEG becomes up to 250\% more search efficient in the 99\% recall regime of the GloVe dataset. Future work will address vertex deletion in order to handle dynamic datasets, improving thread safety for multi-threading purposes, and adding support for filters and multi-modal search.
\newpage
%
%
%
\bibliographystyle{ACM-Reference-Format}
\balance
\makeatletter
\if@ACM@anonymous
    \bibliography{references,references_anon}
\else
    \bibliography{references,references_self}
\fi
\makeatother

\end{document}